Characterizing the structure of mouse behavior
using Motion Sequencing

Sherry Lin*[1], Winthrop F. Gillis*[1], Caleb Weinreb[1], Ayman Zeine[1],
Samuel C. Jones[1], Emma M. Robinson[1], Jeffrey Markowitz[1,2], Sandeep Robert Datta[1,#]

[1]Department of Neurobiology, Harvard Medical School
*These authors contributed equally
[2]Present address: Wallace H. Coulter Department of Biomedical Engineering, Georgia Institute of Technology and Emory University
[#]Corresponding Author

**Abstract**

Spontaneous mouse behavior is composed from repeatedly-used modules of movement (e.g., rearing, running, grooming) that are flexibly placed into sequences whose content evolves over time. By identifying behavioral modules and the order in which they are expressed, researchers can gain insight into the impact of drugs, genes, context, sensory stimuli and neural activity on natural behavior. Here we present a protocol for performing Motion Sequencing (MoSeq), an ethologically-inspired method that uses 3D machine vision and unsupervised machine learning to decompose spontaneous mouse behavior in the laboratory into a series of elemental modules called "syllables". This protocol is based upon a notebook-based pipeline for MoSeq that includes modules for depth video acquisition, data pre-processing and modeling, as well as a standardized set of visualization tools. Users are provided with instructions and code for building a MoSeq imaging rig and acquiring three-dimensional video of spontaneous mouse behavior for submission to the modeling framework; the outputs of this protocol include syllable labels for each frame of the video data as well as summary plots describing how often each syllable was used and how syllables transitioned from one to the other over time. This protocol and the accompanying pipeline significantly lower the bar for adopting this unsupervised, data-driven approach to characterizing mouse behavior, enabling users without significant computational ethology experience to gain insight into how the structure of behavior is altered after experimental manipulations.

**Introduction**

Classical work from ethology argues that animals compose their behaviors, at least in part, from a context-specific set of repeatedly-used, three-dimensional, and stereotyped motifs of movement[1-3]. These behavioral modules are expressed in predictable sequences over time and organized hierarchically, thereby affording behavior structure at multiple timescales. Identifying the behavioral modules and sequences expressed by an animal during an experiment can aid researchers in: better understanding the basic principles that organize action; revealing mechanisms through which genes and neural circuits collaborate to build behavior; characterizing the impact of disease genes or processes on behavior; and developing therapeutic agents aimed at addressing disease phenotypes[4-6].

Behavioral modules have traditionally been identified through careful human observation of animal behavior. Advances in machine vision and machine learning have recently revolutionized methods for behavioral measurement and characterization, both in laboratory settings and in the natural world. For example, supervised machine learning, in which researchers pre-specify the number and content of behavioral modules, has been successfully used to automatically identify a wide variety of behavioral modules during the spontaneous behavior of flies, worms and mice[7-15]. Unsupervised approaches that discover behavioral modules in a data-driven manner have arguably been less well developed (see Comparisons to other methods, below), but promise to unveil key aspects of animal behavior that may be missed by conventional human observation[16-23].

Our lab has articulated an unsupervised framework to characterize behavior called Motion Sequencing (MoSeq), which can identify behavioral modules expressed by mice in the laboratory[21,22,24-27]. MoSeq takes as its inputs three-dimensional (3D) depth videos of mice behaving in simple arenas, models these data using statistical learning and Bayesian non-parametric approaches, and then outputs a set of behavioral modules — which we call "syllables" — that describes the behavior observed in the experiment. While MoSeq is a powerful approach to behavioral phenotyping, its adoption has been significantly limited by the complexity of the underlying code, and by the attendant inaccessibility to researchers who lack expertise in machine learning.

Here we present a user-friendly and validated MoSeq protocol that is based upon a newly-developed pipeline that can be deployed by users with little prior experience in computational ethology. We include instructions on how to build a MoSeq rig capable of acquiring 3D depth videos, and for using the python-based MoSeq pipeline to model and analyze the behavioral imaging data. We also include instructions for running a modeling and analysis pipeline that generates interpretable outputs describing the behavioral modules and sequences identified from the acquired depth videos. Although a detailed technical description of MoSeq is beyond the purview of this protocol (see[21] for a detailed description of the underlying methodology), in brief this pipeline identifies, extracts and aligns the depth image of each mouse along the axis of its spine, dimensionally reduces these cropped and aligned mouse images, and then models the dimensionally-reduced time-series data using an autoregressive sticky Hidden Markov Model (AR-HMM). The AR-HMM is fit using Gibbs sampling, using a hierarchical Dirichlet prior over the number of syllables, enabling MoSeq to automatically learn the number of behavioral syllables without user pre-specification. The output of this procedure is a sequence of labels indicating the most likely behavioral syllable expressed during each frame of the behavioral video, and a kinematic summary of each syllable (which is described as an autoregressive process through pose space). The MoSeq pipeline also provides tools to extract conventional behavioral parameters from the data (e.g., velocity or position plots), to visualize syllable examples as movies, to plot how often each syllable is used in the experiment, and to plot transition statistics between syllables. The presented MoSeq protocol thereby enables users to decompose mouse behavior into its constituent modules and sequences, and to quantitatively analyze how spontaneous mouse behavior might be similar or different across a variety of experimental conditions.

Applications

MoSeq has been used to characterize mouse behavior in a variety of experimental contexts (Fig. 1). Wiltschko et al, in the initial development of MoSeq, demonstrated that MoSeq captures behavioral differences attributable to changes in the physical structure of the environment, salient sensory cues, and both genetic and optogenetic manipulations[21]; similar experiments by other labs have used MoSeq to capture the behavioral effects of genetic and neural perturbations[26,28-30]. In addition, MoSeq has been successfully used to characterize the effects of pharmacological agents on behavior (and to build systematic maps of drug space based upon mouse behavior), to identify neural correlates of syllables and sequence statistics in the striatum, and to reveal the effects of brain lesions on patterned action[22,25]. Other use cases can be found on the MoSeq website (http://www.MoSeq4all.org). In this protocol we include data representing one typical use case, in which mice are administered different experimental treatments (in this case, amphetamine and saline) expected to elicit different patterns of behavior.

Advantages

The MoSeq AR-HMM Model:
Most methods for unsupervised behavior segmentation involve transforming raw data using local dynamical features and then clustering timepoints within the resulting feature space or within a low-dimensional embedding of that space. The model-based approach at the core of MoSeq confers several unique advantages. First, because every frame represents a distinct state in a Markov chain, behavioral boundaries are localized to "changepoints", facilitating analyses of behavior timing. Second, by explicitly modeling the transition probabilities between behavioral states, the AR-HMM incorporates temporal context in the classification of each frame. Third, MoSeq uses non-parametric Bayes to automatically determine the number of distinct behavior states, adaptively adjusting the level of coarse-graining based on the amount of data and the inherent complexity of the observed behaviors.

Extraction and Modeling pipeline:

Reaping the full benefits of model-based analysis requires an initial investment in data quality and proper data processing. A key advantage of the MoSeq analysis pipeline described herein is its focus on data curation and quality control. Various steps in the pipeline include tools (like online visualizers) that enable users to flag potential problems and facilitate rapid iteration. A reference dataset is also provided to enable new users to quickly judge the relative quality of their data and the model outputs.

Visualization pipeline:
Whereas clustering-based behavioral analysis can be easily visualized, e.g., using scatter plots of behavioral embeddings, the outputs of MoSeq AR-HMM models – such as syllable boundaries and transition probabilities – are more abstract. An advantage of the new analysis and visualization pipeline is the wealth of different plotting/visualization modalities that can now be accessed with a few keystrokes. Users can now easily visualize the distribution of syllable usages while displaying movies of mice performing the syllables. It is also simple to generate graph-based visualizations of syllable transitions, and to depict how the transition statistics shift between experiments.

Limitations

There are several limitations in the MoSeq framework. The AR-HMM is hierarchical — syllables are described by the model as continuous auto-regressive trajectories through pose space, whose transitions are specified by a sticky HMM. This means that MoSeq is explicitly seeking to explain behavior in situations in which pose dynamics are continuous over sub-second timescales, with different patterns of movement switching from one to another as behavior evolves. It is important to note that a simple inspection of raw depth mouse behavioral data reveals that it is inherently broken up into sub-second blocks that appear to correspond to behavioral modules, suggesting that MoSeq should effectively capture key aspects of mouse behavior in many situations[21]; furthermore, we have previously demonstrated that the behavioral dynamics captured by MoSeq have explicit correlates in neural centers responsible for action selection and sequencing[25]. However, MoSeq may not be well suited for capturing structure in other types of behavioral data that lack the continuous-discrete dynamics instantiated by the AR-HMM.

One goal of the MoSeq pipeline is to discover syllables and syllable relationships in an unsupervised manner. This necessarily means that the underlying model requires retraining each time new data is acquired. Any datasets that a user wishes to directly compare must be co-modeled, rather than modeled separately and then compared post hoc, as syllable labels are not shared across models. MoSeq also characterizes behavior at a single timescale, which is influenced by a hyperparameter available for manipulation in the pipeline called kappa; kappa roughly specifies how "sticky" each syllable is in the AR-HMM, and therefore influences the overall duration distribution of syllables identified by MoSeq. We provide tools in the MoSeq pipeline to set the kappa parameter based upon the block structure that is inherent in the depth data. One important consequence of using an AR-HMM framework is that each moment in time is assigned one and only one syllable label. Thus, MoSeq characterizes moments during which mice are doing two behaviors at once (for example, walking and turning) as a single unified behavior (for example, walking while turning) rather than as two separate behaviors that coincide in time.

The MoSeq pipeline is sensitive to mouse morphology and size, requiring all mice in an experiment to be of roughly similar proportions. The model is also sensitive to the amount of data to which it is applied, with the number of identified syllables scaling sublinearly with data size. Theoretically, the model underlying MoSeq should have the capability to generalize to other data sources (e.g., point-tracking data or 2D videos), multiple animals and more complex experimental environments. However, the framework we describe at present only supports analysis of individual animals exploring simple environments, such as open fields, elevated plus and Y mazes. The depth cameras currently used by the MoSeq pipeline capture pose dynamics in roughly the top two-thirds of the mouse carapace, as the depth cameras used in this protocol are placed above the arena, and therefore observe mouse behavior from the top down. As limb movements cause systematic distortions in the shoulder and hip girdles, many behaviors associated with differential limb movements (i.e., walking vs running) are easily distinguished by Moseq; that said, limb movements and gait details per se are not captured using the current "camera from above" approach.

Data acquisition requires specialized hardware. The MoSeq pipeline was designed to work specifically with the Kinect v2 depth camera. The camera must be set up within a specific height range from the floor and the recording environment requires regular maintenance such as cleaning, sanding, and painting to ensure the quality of the recordings. Finally, data acquisition requires significant storage space. For example, an uncompressed 20-minute session requires ~20 GB on a hard drive.

Comparisons to other methods

It is important to distinguish the purpose and function of MoSeq (which at a high level reveals the underlying structure of behavior) from pose estimation methods, which uses deep learning approaches to track the position of user-identified keypoints in behavioral video[31-36]. Furthermore, MoSeq is an unsupervised behavioral analysis technique, meaning that it learns the number and identity of behavioral syllables in any given experiment from its intrinsic spatiotemporal structure; this is unlike supervised machine vision/learning methods, which identify user-specified behaviors based upon ground truth training data[5,27,37]. One important additional difference between MoSeq and supervised methods is that MoSeq assigns a unique syllable label to each frame, whereas multiple supervised behavioral classifiers can be run in parallel over video data, enabling identification of multiple behaviors that co-occur at the same time.

There are two main alternative unsupervised machine learning-based behavioral classifiers in current use. The first approach is MotionMapper, which takes two-dimensional behavioral videos, re-featurizes those videos based upon spectral content, and then embeds each re-featurized video frame into a two-dimensional space using t-stochastic nearest neighbors (t-SNE) embedding[16,38-41]. This approach differs from MoSeq in that the data are not subject to time-series modeling (which in an AR-HMM necessarily focuses on a single timescale), but instead behavioral modules are identified via watershedding of the t-SNE embedding. The similarities and differences between MoSeq and MotionMapper (and their relative advantages and disadvantages) have been previously considered[27]; in brief, the unitary timescale of MoSeq affords significant advantages for characterizing the sub-second structure of behavior, but at a potential cost of losing explicit information about longer timescale hierarchical structure that is maintained by MotionMapper.

The second approach is b-SOiD, which takes as its inputs keypoints derived from pose estimation methods (e.g., DeepLabCut), embeds these data after featurization into a low-dimensional space using Uniform Manifold Approximation and Projection (UMAP), and then performs hierarchical clustering to identify behavioral modules[20]. B-SOiD is similar to MotionMapper insofar as it does not subject the data to time-series modeling. Furthermore, the dimensionality of the input data to b-SOiD is typically lower than that obtained from either two-dimensional or depth cameras; using low dimensional descriptions of behavior has empirically been shown to reduce performance in behavioral classification tasks[22]. However, both MotionMapper and b-SOiD can be applied to standard two-dimensional behavioral videos, whereas the MoSeq framework in its current form is limited to depth video. Note that there is nothing in the MoSeq framework that prevents time-series modeling of keypoints or related data, and we anticipate simple extensions to the MoSeq framework in the near future to enable this capability.

Protocol development

In its initial form as reported in Wiltschko et al[21], key aspects of MoSeq were distributed across a variety of independent codebases, each of which had its own dependencies and compute requirements. To convert these distributed codebases into a unitary plug-and-play pipeline that can be incorporated into a standardized protocol, we addressed several key challenges:

1. We formalized procedures for constructing a MoSeq depth camera setup and established standards for data acquisition that ensures quality modeling in downstream steps.
2. We created a modular MoSeq pipeline by refactoring the underlying MoSeq framework into a Jupyter notebook format. This pipeline serially carries mouse depth imaging data through each of several key steps, including mouse image extraction, alignment and quality control, data modeling (with the ability to flexibly change key hyperparameters), and data visualization.
3. We generated both Anaconda and Docker-based installation routes to simplify installation and to avoid missing dependencies or code conflicts.

4. We formalized the code used for data analysis and visualization, which allows a large suite of informative metrics to be automatically generated at the end of each experiment.
5. We generated online training resources, including how-to-videos and detailed documentation to supplement written protocols (see http://www.MoSeq4all.org).

Overview of the protocol

Code access, instruction videos, access to community resources and updates are provided via the MoSeq website (http://www.MoSeq4all.org).

The MoSeq pipeline and this protocol consists of three procedures: 1) acquisition of depth video, 2) extraction and modeling, and 3) analysis and visualization (Fig. 2).

Depth video acquisition:
Procedure 1 describes the process of acquiring depth video (Fig. 3). First, session metadata is added to the appropriate fields for a given experiment. Then, a recording configuration is defined, and the depth recording of mouse behavior is performed. Finally, the data is organized into a unique folder for each session.

Extraction and modeling:
Procedure 2 outlines the process of extracting the image of the mouse from depth videos (Fig. 4) and fitting an AR-HMM model to these extracted data to identify the syllables expressed by the mouse during the experiment. First, depth image extraction parameters are set by using an interactive widget, which enables the pipeline to crop, center and orient the image of the mouse from the depth data; this procedure uses a flip classifier to ensure that after this procedure the nose of the mouse in the images is always pointing to the right. Then, kinematic parameters such as velocity, height, and position are computed. Users then label each imaging session and assign it to a particular experimental group (e.g., control versus drug-treated, Fig. 5), after which the extracted videos are dimensionally reduced (Fig. 6) and modeled using an AR-HMM (Fig. 7).

Analysis and visualization
Procedure 3 describes the steps taken to visualize and interpret the output of MoSeq. Users assign labels to individual MoSeq-identified syllables (e.g., rearing, running, sniffing) by watching MoSeq-generated "crowd" movies and by using a provided labeling tool (Fig. 8a). Similarities and differences between experimental groups are visualized by plotting syllable usage distributions (Fig. 8b) and transition graphs (Fig. 8c).

**Materials**

Behavioral acquisition hardware
Acquisition computer: a desktop computer with the following minimum specifications: i5-6400 - 2.7GHz (up to 3.3GHz) Intel Quad Core - 6MB Cache, HD Graphics 530 (65W), at least 1 USB 3.0 port (Gen 2 or above), 8GB RAM, 1.0TB of storage, a PCIe slot for the GPU (see next item), and a power supply sufficient to power both the computer and GPU. Note that the acquisition computer is not the same computer used to run the MoSeq analysis pipeline.

CRITICAL: A large amount of storage is necessary as a single 20-minute, uncompressed depth video is ~20 GB in size.

GPU: a GPU with DirectX 11 support is necessary to use the acquisition software. Many Nvidia GPUs are DirectX 11 compatible, such as the GeForce GTX 960 or the RTX 20 and 30 series.

Microsoft Kinect v2 depth camera: a depth-sensing camera for recording 3D depth videos.

CRITICAL: The Kinect v2 is the only camera currently supported by the pipeline, but support for additional cameras (including the Microsoft Azure and the Intel Realsense) will be included in future software releases; see http://www.MoSeq4all.org for updates on camera compatibility).

MoSeq Recording cage: The recording cage is a rigid cage constructed using 80/20 aluminum bars. The arena sits in the center and the depth camera is mounted on the top. Table 1 is the bill of materials (https://bit.ly/3tHNU3p) specifies the exact materials used to construct the cage.

Black polyethylene bucket: the standard arena for the mouse is a 14" tall black polyethylene bucket. The bucket must be sanded and painted with matte black paint to reduce reflections to improve recording quality. Table 1 specifies the source for the bucket, sandpaper and the matte black paint used to prepare the bucket. Arenas in other custom configurations (i.e., elevated plus mazes, Y mazes) can be assembled using acrylic sheets painted with matte black paint.

CRITICAL: Reflections from the infrared source used by the Kinect v2 camera are a major source of data degradation; only use properly prepared buckets/arenas for experiments using this camera. Using the recording cage to mount the camera is strongly recommended, as it standardizes the camera height and in many cases prevents the transmission of vibrations (which are pervasive when using e.g., boom arms).

Acquisition Software

Note: Installation instructions for this required code are described in the Equipment Setup and Software Installation section below.

Operating system: We have tested the acquisition software on Windows 10. Other Windows versions may be compatible, but the software has not been tested in alternative environments.

Kinect for Windows Runtime 2.0 (https://www.microsoft.com/en-us/download/details.aspx?id=44559): The software that provides the required drivers and runtime environment for Windows applications to use Kinect sensor technology.

Kinect for Windows SDK 2.0 (https://www.microsoft.com/en-us/download/details.aspx?id=44561): The software development kit for developers to use Kinect2 sensor technology.

Ni-DAQmx version 17.6 (http://sine.ni.com/nips/cds/view/p/lang/en/nid/10181): National Instruments (NI) software for auxiliary data acquisition.

CRITICAL: Even if NI devices are not used, Ni-DAQmx is required as part of kinect-nidaq.exe installation.

The kinect2-nidaq GitHub repository: contains the executable kinect2_nidaq.exe version 0.2.3, the acquisition graphical user interface (GUI) for acquiring depth videos using Kinect2. Please visit the MoSeq website (http://www.MoSeq4all.org) and follow the instructions to request access to this repository. Note that obtaining all MoSeq2 software components hosted on GitHub — including access to the MoSeq2 wiki — requires first obtaining access permissions via the MoSeq website.

Data Analysis Hardware

Analysis computer: We recommend running MoSeq analysis on a computer with at least 8 cores and 32GB of RAM. The number of cores will dictate the speed of the model training step, as the model is parallelized across cores. MoSeq analysis can also be run on a high-performance computing cluster that utilizes Slurm (i.e., university research computing, Amazon Web Services, or the Google Could Platform) to parallelize the extraction, dimensionality reduction, and modeling steps.

Software for Analysis Computer

Note: Installation instructions for this required code are described in the Equipment Setup and Software Installation section below.

Operating system: Linux (Tested on Ubuntu 16.04 LTS or higher and CentOS), Windows 10 with Windows Subsystem for Linux version 2 (WSL2) with Ubuntu Kernel (https://docs.microsoft.com/en-us/windows/wsl/install), or macOS Version 11 Big Sur or higher. At the time of this writing, the use of MacBooks with the M1 chip is not supported; in this case, we recommend installing MoSeq in a remote computing environment using Linux.

Anaconda (https://www.anaconda.com/): a distribution of Python for scientific computing with simple management for Python packages and virtual environments. The MoSeq2 package suite is written using Python 3.7 and is not compatible with Python 2.

Git (https://git-scm.com/): A version control system that tracks content changes. MoSeq2 package suite is hosted on GitHub and uses Git to facilitate version control and download.

Curl (https://curl.se/): A command-line tool for networked data transfer.

The gcc and g++ compiler: MoSeq2 uses C-accelerated python code to facilitate the AR-HMM training process, therefore the gcc and g++ compilers must be installed before installing the MoSeq2 package suite. Versions 6, 7, 9, or 10 are supported.

The latest version of moseq2-app (version 1.1.2 at the time of writing): the MoSeq2 package suite covers the entire MoSeq pipeline via interactive Jupyter notebooks; it also includes a version of MoSeq that can be run through the command-line interface (CLI). Please visit the MoSeq website (http://www.MoSeq4all.org) and follow the instructions to request access to the package suite.

The toolbox includes:
moseq2-extract (version 1.1.2 at the time of writing)
moseq2-pca (version 1.1.3 at the time of writing)
moseq2-model (version 1.1.2 at the time of writing)
moseq2-viz (version 1.2.2 at the time of writing)
moseq2-app (version 1.2.3 at the time of writing)

Jupyter notebooks (https://jupyter.org/install): We provide two Jupyter notebooks to run the MoSeq pipeline as part of the MoSeq package: a notebook for extraction and modeling, and a notebook for analysis and visualization of the results.

Docker Desktop (https://docs.docker.com/get-started/) and MoSeq Docker Image (https://hub.docker.com/repository/docker/dattalab/moseq2): Docker is supported on Linux, macOS, and Windows 10 through WSL2. The MoSeq Docker image, obtainable from the moseq2 docker hub repository (https://hub.docker.com/repository/docker/dattalab/moseq2) is a containerized version of the MoSeq2 package suite that runs using Docker Desktop. The Docker image comes with the MoSeq packages, the MoSeq2 Jupyter notebooks, and CLI tools pre-installed. The MoSeq Docker image is intended for users that are new to programming and not familiar with managing packages and virtual environments with Anaconda. It is sufficient to install either the MoSeq2 package suite directly in virtual environments using Anaconda or using the MoSeq Docker image; it is not necessary to have both for the MoSeq2 package suite to work correctly.

Example MoSeq dataset (https://github.com/dattalab/moseq2-app/wiki/Download-Test-Data): we provide two sample datasets that can be downloaded from the GitHub wiki as part of the MoSeq package. The download scripts are also available in the scripts folder, which is located within the moseq2-app folder. Each dataset contains depth video recordings of C57 mice injected with either saline (control group) or amphetamine. One dataset has a total of 48 recordings, and the other includes 20 recordings. Each dataset comes with pre-trained models for users to test and familiarize themselves with the MoSeq analysis pipeline; these datasets are required to obtain the expected results shown herein.

**Equipment Setup and Software Installation**

Acquisition Hardware setup

The MoSeq acquisition apparatus includes a recording cage, depth camera, and arena. In this protocol, the arena is a sanded and painted plastic bucket.

1. Assemble the recording cage and mount the camera. Timing ~2 hours

    Follow the visual instructions for assembling the cage on the GitHub wiki (https://github.com/dattalab/moseq2-app/wiki/MoSeq-Cage). The depth camera is mounted on the cage and the camera must be positioned directly in the center above the area to accurately capture all the depth values (Fig. 3b).

CRITICAL: The mounted depth camera should be parallel to the floor at a height of 673 ± 5 mm (or 26.5 in). Incorrect camera height can lead to data loss. When the camera is too far from the floor of the arena, the mouse is composed of fewer pixels (thereby reducing the resolution MoSeq has over behavioral syllables), whereas when the camera is too close, part of the mouse might be clipped if the mouse exceeds the minimum distance threshold of the camera.

2. Prepare the bucket. Timing ~20 minutes

    Sand the bucket using 80-grit sandpaper. The original smooth surface on the base and 15 cm up along the walls should be completely sanded over, and no smooth or reflective areas should remain. Any patches of smooth surfaces will cause infrared reflections, which degrade the quality of the acquired Kinect data (Fig. 3c). Sand the base of the bucket firmly with consistent ~2" circular motions while removing the plastic debris. After the base is completely sanded, lay the bucket down on its side and sand the lower 15 cm of the bucket wall using the same circular motion.

CRITICAL: Ensure that the surface is uniformly sanded and that there are no reflective smooth spots, especially the wall-floor junction. Failure to do so will cause reflections and image degradation.

3. Paint the open field. Timing ~20 minutes for painting and ~24 hours for drying.

Spray a modest coat of the Krylon flat-black paint (listed in Table 1) on the inner walls of the bucket and let dry and overnight. The bucket can be scrubbed 24 hours after painting with a soft cloth or sponge with soap and warm water to reduce the lingering paint smell.

Acquisition software installation

Follow the instructions below to install acquisition software on the acquisition computer. A more detailed description can be found on the acquisition software wiki (https://github.com/dattalab/kinect2-nidaq/wiki/Acquiring-data---Kinect-v2).

4. Download kinect2-nidaq version 0.2.3 from GitHub (https://github.com/dattalab/kinect2-nidaq). Unzip the contents onto the Desktop. Timing ~15 minutes

5. Download and install the Kinect for Windows Runtime 2.0 (https://www.microsoft.com/en-us/download/details.aspx?id=44559) Timing ~30 minutes

6. Download and install the Kinect for Windows SDK 2.0 (https://www.microsoft.com/en-us/download/details.aspx?id=44561) Timing ~30 minutes

7. Download and Install version 17.6 of NI-DAQmx (http://sine.ni.com/nips/cds/view/p/lang/en/nid/10181), even if a National Instruments device is not used. Timing ~1 hour
    a. Select 17.6 from the Version dropdown menu and click "DOWNLOAD" to download the installer. Note that a National Instruments account is required for the download.
    b. Install NI-DAQmx from the installer.

CRITICAL: During the installation, make sure to include support for the .NET Framework 4.0, 4.5, and 4.5.1 Languages Support.

MoSeq2 package suite installation

The MoSeq2 package suite has two installation routes: either via Docker or Anaconda. For users less familiar with scientific computing, we strongly recommend installing the MoSeq2 package suite with Docker. Note that using the MoSeq2 package suite from the Docker container reduces performance, and so is not recommended for large datasets. In these circumstances, performing MoSeq using a high-performance computing cluster and the Linux operating system is recommended, which requires installing the MoSeq2 package suite through Anaconda. Details for installing MoSeq via the MoSeq Docker image or via Conda can be found in moseq2-app wiki (https://github.com/dattalab/moseq2-app/wiki).

8. Install the MoSeq2 package suite via Docker. Timing ~1 hour

a. Install Docker Desktop and create an account with Docker Hub (https://hub.docker.com/signup). One can find more information about Docker on the official Docker website (https://docs.docker.com/get-started/).

b. Launch Docker Desktop and sign in with a Docker ID.

c. Set up a folder in a convenient location on the analysis computer that will store MoSeq-related data, such as a folder called "example_data" as the specific project directory.

d. Navigate to the REMOTE REPOSITORIES tab in Images and select "dattalab" from the dropdown menu.

e. Pull (download) the MoSeq2 Docker image from the dattalab MoSeq2 repository.

f. Navigate to the LOCAL tab in Images and click RUN to use the MoSeq2 package suite in a Docker container.

g. Click "Optional settings" to expand the container configuration options and input the following settings into the respective text fields:
   Host port: 8888
   Host path: This field is used to indicate the path to the specific project directory. The … button to the left can be used to select the folder.
   Container path: /data

h. Click the RUN button to start the container.

Up-to-date configuration and video examples of configuring and running the MoSeq2 package suite in a Docker container is described in detail in the wiki (https://github.com/dattalab/moseq2-app/wiki#installation-via-docker).

8. (Alternative) Install the MoSeq2 package suite using Anaconda. Timing ~30 minutes

All the following commands should be run in a command-line interpreter such as Terminal in macOS or Linux. For Windows, run the commands in the Ubuntu terminal through WSL2.

a. Set up Git credential storage. Inputting the GitHub credentials is required multiple times during the installation process.

Remember GitHub credentials by typing the following command:

>>> git config --global credential.helper cache

b. If you are using MacOS, install gcc version 7 using homebrew (https://brew.sh/) with the following command into the Terminal:
>> brew install gcc@7

c. Export gcc and g++ paths by typing the following command:
In MacOS:
>>> export CC="$(which gcc-7)"
>>> export CXX="$(which g++-7)"
In WSL2 or Linux:
>>> export CC="$(which gcc)"
>>> export CXX="$(which g++)"

CRITICAL: Make sure gcc is at least version 6 or higher, which can be found by typing the following command:
>>> gcc --version

d. Clone the moseq2-app GitHub repository into a new folder by typing the following command:
>>> git clone -b release https://github.com/dattalab/moseq2-app.git

Input your username and password for GitHub when prompted.

CRITICAL: If two-factor authentication is enabled, instead of your GitHub password, a personal token for downloading repositories will be used in place of the password. Follow the official documentation (https://docs.github.com/en/authentication/securing-your-account-with-two-factor-authentication-2fa/accessing-github-using-two-factor-authentication) on GitHub to generate a personal token.

e. Navigate into the moseq2-app folder and create a conda environment named moseq2-app using the moseq2-env.yaml config file stored in the scripts directory by entering the following commands:
>>> cd moseq2-app
>>> conda config --set channel_priority strict
>>> conda env create -n moseq2-app --file scripts/moseq2-env.yaml

f. Activate the newly created conda environment by entering the following command:
>>> conda activate moseq2-app

g. Install all the packages in the MoSeq2 package suite using the installation script stored in the scripts folder, by entering the following command:
>>> ./scripts/install_moseq2_app.sh

h. Check the versions of all packages to verify they were installed correctly by entering the following commands:
>>> moseq2-extract --version
>>> moseq2-pca --version
>>> moseq2-model --version
>>> moseq2-viz --version

**Procedure 1: Data Acquisition**

1.  Enter parameters and metadata for imaging. Timing ~10 minutes

    a. Click on the kinect2-nidaq.exe application to launch the acquisition GUI (Fig. 3d).

    b. Preview the recording by checking the "Preview Mode" checkbox (Fig. 3d, label 1) and clicking the "Start Session" button (Fig. 3d, label 2). Go to the "Preview" tab (Fig. 3d, label 3) to see both the RGB and depth videos. Adjust the two sliders indicating the minimum and maximum distance threshold as is appropriate given the positioning of the arena and camera.

    c. Input recording metadata in the text input boxes.

    > Session Name (Fig. 3d, label 4): This field is used to indicate the date, cohort, experimental condition and/or environment type.

    > Subject Name (Fig. 3d, label 5): This field is used to indicate the rodent strain, sex, age and/or additional identifiers. The subject name should uniquely identify each mouse.

    > Save Directory (Fig. 3d, label 6): This field is used to indicate the path to save the acquired data. The … button to the left can be used to select a folder to save the data.

    d. Indicate the types of data to be acquired during the experiment. Check the "Depth Stream" checkbox (Fig. 3d, label 7 top) to enable the acquisition of depth data. Check the "Color Stream" checkbox (Fig. 3d, label 7 bottom) to enable the acquisition of the RGB stream. Color video is not used for any downstream analysis in the MoSeq pipeline, but can be useful for other behavioral analyses.

CRITICAL: Only record RGB data (in addition to the depth data) if necessary. Recording both depth and RGB streams increases storage requirements, and depending upon system specifics, can lead to a high number of dropped frames.

    e. (Optional) Check the "Compress Session" checkbox to compress the recording (Fig. 3d label 8). A gzipped tarball with a timestamp will be created at the end of the recording, which contains the files described in step 2c.

    f. Indicate the desired imaging time. Type in the time you wish to record in the "Recording time (minutes)" field (Fig. 3d, label 9). A typical mouse behavioral recording in the open field lasts for 25 to 30 minutes. The video recording will automatically end after the specified number of minutes. Alternatively, check the "Record until user clicks stop" box (Fig. 3d, label 10) to continuously record video until the "Stop Session" button (Fig. 3d, label 11) is clicked.

CRITICAL: Mouse behavior is decidedly non-stationary, with behavior in the first five minutes of the recording differing sharply from the next five minutes, and so on. We recommend imaging in the open field for 25-30 minutes because at the end of that time mice often habituate to the arena and stop moving. Note that the minimum effective data size for modeling is ~300 total minutes (~1 million frames at 30 fps), so choose per-session imaging times and the total number of mice in a given experiment accordingly. We typically use 8-10 mice per experimental condition, each of which is imaged for 30 minutes; thus, depending upon the number of experimental conditions, the amount of data submitted for modeling often range from 2 to 20 million frames.

2. Image mouse behavior. Timing ~30 minutes

a. Gently place the mouse into the arena.

b. Start imaging by clicking the "Start Session" button (Fig. 3d, label 2). The recording will automatically stop when the time for the preset recording time is over or when the "Stop Session" button (Fig. 3d, label 11) is clicked.

c. Check to ensure the files were recorded and in the appropriate location on your hard drive.

   When the recording is done, the following files are created in a folder called session_<datetime>, for example, "session_20220526103416". Files in this session-specific folder will include:

   - depth_ts.txt: The depth_ts.txt file records the timestamps of each video frame in plain text format. The file has 2 columns separated by a single whitespace. The first column contains hardware timestamps of the camera in milliseconds while the second column contains timestamps from the NiDAQ if NiDAQ data capture is used. If NiDAQ data capture is not used, the second column will be populated with zeros. The MoSeq2 package suite only uses the first column.

   - depth.dat: The depth.dat file is a 3D depth video stored in raw byte form. Each pixel of each movie frame is a little-endian unsigned 16-bit integer (uint16) representing the distance from the camera, in millimeters.

   - metadata.json: The metadata.json file contains the following information in JSON format: mouse name, session name, time of the recording, NIDAQ-specific parameters, and video-specific parameters.

d. Clean the bucket after each recording session. Spray the arena with a 10% bleach solution and wipe dry with a paper towel. Then, spray the arena with a 70% ethanol solution and wipe it dry.

3. Transfer data to the analysis computer. All session subfolders should be in one project folder. Timing variable

**Procedure 2: MoSeq extraction and modeling**

Each MoSeq project is contained within a project directory. The project directory is defined as the folder that contains the relevant depth videos (i.e., "sessions") and will ultimately house the output from the MoSeq pipeline. At the beginning of the project, the project directory should have separate subfolders for each depth recording session (Table 2.1). The directory file structure at the end of each step is described in detail in the GitHub wiki (https://github.com/dattalab/moseq2-app/wiki/Directory-Structures-and-yaml-Files-in-MoSeq-Pipeline).

Data can be extracted and modeled using either a Jupyter Notebook or using the command line interface (CLI). This protocol focuses on using the Jupyter Notebook for extracting and modeling the data; a detailed description of the CLI alternatives can be found in the GitHub wiki (https://github.com/dattalab/moseq2-app/wiki/Command-Line-Interface-for-Extraction-and-Modeling).

To better organize the results, we recommend copying all the MoSeq2 Jupyter Notebooks from the moseq2-app/notebooks folder in the moseq2-app GitHub repository to the project directory. In this protocol, the MoSeq2 Jupyter Notebooks are assumed to have been copied to the project directory.

1. Start the MoSeq2 Extract Modeling Jupyter Notebook. Timing ~2 minutes

CRITICAL: Anaconda users run steps a-f; Docker users run step g-h.

   a. (Anaconda users only) Open Terminal on the analysis computer. Terminal is typically available in the applications directory.

   b. (Anaconda users only) Activate the MoSeq Conda environment by entering the following command, assuming the name of the environment is moseq2-app:
      >>> conda activate moseq2-app

   c. (Anaconda users only) Navigate to the project directory in Terminal. To do so, enter the following command (replace desktop/project with the path to the specific project directory):
      >>> cd desktop/project

   d. (Anaconda users only) Check the contents in the project directory by opening the directory in the finder, or by entering the following command in Terminal:
      >>> ls

   Folder contents should be organized as indicated in Table 2.1

   e. (Anaconda users only) Launch the Jupyter server by entering the following command in Terminal:
      >>> jupyter notebook

This will open up a webpage that indicates all of the contents in the project directory.

f. (Anaconda users only) Launch the MoSeq notebook by clicking "MoSeq2-Extract-Modeling-Notebook.ipynb" This will open the MoSeq2 notebook in a new tab.

g. (Docker users only) Once the Docker container is running, click the "OPEN WITH BROWSER" icon to open a webpage to run the Jupyter notebook. Input "moseq" in the password text field to display all the content in the project directory.

h. (Docker users only) Launch the MoSeq notebook by clicking "MoSeq2-Extract-Modeling-Notebook.ipynb" This will open the MoSeq2 notebook in a new tab.

2. Validate and initialize the project. Timing ~5 minutes

a. Validate that the correct MoSeq packages are installed by running the first code cell. Here and throughout, to run a Jupyter notebook cell first click into the cell, and then type shift+enter. You will know a given cell is running when the "In" icon is associated with a * symbol. When the * is converted into a number, the cell has finished running (Fig. 9a).

Initialize or restore progress variables by running the Set Up or Restore Progress Variables cell. This cell must be run every time the notebook is used and every time the progress.yaml file is changed outside of the notebook. Progress variables — which keep track of where metadata are stored, and which monitor progress along the MoSeq pipeline during analysis — are stored in a progress.yaml file (whose contents are described in Table 3), which lives in the project folder, specified in base_dir variable field ("./" indicates that the project folder is the folder the notebooks are in). If this is the first time the analysis pipeline is being run on a particular project, you will initialize the progress variables and generate a progress.yaml file by simply running this cell. To restore previous progress variables from prior or interrupted analyses, you can use this cell to indicate which progress.yaml file you wish to use. To load progress variables different from those saved in the progress.yaml in the project folder, simply replace the current progress.yaml file that lives in the project folder with the one you wish to be used.

b. Run the Generate Configuration File cell to generate the config.yaml file. This cell generates a config.yaml that holds all configurable parameters for all steps in the MoSeq pipeline, such as those related to extraction and principal components analysis (PCA). This file is initialized with default values that have been found to work well for extracting depth data of adult C57BL/6J mice. Table 4 shows the key parameters saved in config.yaml file.

c. Run the Download a Pre-trained Flip Classifier Model File cell, which enables users to download a flip classifier for image processing. Flip classifiers are random forest classifiers that take as their input mouse images, and output whether the mouse nose is pointing to the right or the left in a given depth video frame. The step at which mouse images are extracted from the raw depth video uses the flip classifier to guarantee that the mouse nose is always oriented to the right after cropping and alignment (i.e., to ensure the image of the mouse does not flip from right to left across frames). The flip classifiers are trained for experiments previously run with C57BL/6 mice using Kinect v2 depth cameras. Select a flip classifier type within the Download a Pre-trained Flip Classifier Model File cell to download a Pre-trained Flip Classifier Model File. If the pre-trained flip classifiers fail to correctly identify the mouse's direction in a given dataset, users can take advantage of a flip-classifier training notebook (Flip-Classifier-Training-Notebook.ipynb) that can be found in the same base folder as the MoSeq2 app itself. After using this notebook, add the path of the custom classifier to the flip_classifier field in the config.yaml file.

CRITICAL: Appropriate modeling requires that mice are correctly aligned. Check and change the parameters when necessary to ensure correct alignment; if changing parameters doesn't correct the failed alignment, you must use the custom flip classifier notebook to correctly align your data prior to modeling or exclude the failed sessions.

3. Extract and organize images of the mouse from the raw depth imaging. Timing ~ variable, each session takes ~1 minute

    a. Set session-specific parameters for extracting the image of the mouse by running the Interactive Arena Detection Tool cell (Fig. 4b). To extract the image of the mouse, MoSeq2 identifies the arena (based upon the height of the floor) and segments the image of the mouse away from the arena; the computer vision tools used to perform this extraction require users to set parameters relevant to identifying the floor and the mouse. The interactive tool enables users to visualize how changing these parameters influences the identification of the mouse and the arena (Fig 4b, label 2). Click the Compute arena mask button (Fig 4b, label 5) to apply any parameter changes and compute the mask each time the parameters are adjusted. Similarly, click Compute Extraction (Fig 4b, label 10) to compute a test extraction.

    b. Save the parameters into session_config.yaml file for the extraction step with the "Save session parameters" button (Fig 4b, label 12) once a set of parameters produces a good arena mask and mouse extraction (Fig. 4c and 4d). Alternatively, a user can click "Save session parameter and move to next" button (Fig 4b, label 13) to save the parameters and move to the next session. This button will automatically compute the arena mask and extraction for the next session. This session_config.yaml will be saved in the project folder. The most common parameters are displayed, with more advanced parameters available by clicking the "Show advanced arena mask

parameters" checkbox (Fig 4b, label 6 and label 11). Table 4.1 shows the key parameters used by the tool. If the Interactive Arena Detection Tool is not run, the default parameters in config.yaml file will be used in the extraction step.

    c. Repeat steps 3a and 3b for each session in the project folder.

CRITICAL: If the arena mask is not properly set (Fig. 4d), the mouse will not be correctly identified and extracted (Fig. 4c). All downstream steps that include failed extractions will be corrupted.

CRITICAL: selecting a different session using the session selector (Fig. 4b, label 1) does not automatically compute the arena mask or extraction. After a new session is selected, click Compute arena mask or Compute extraction to compute the arena mask and extraction for the newly selected session.

    d. Run the Extract Session(s) cell to extract mouse images from depth video sessions. The extraction step processes the depth videos to extract the centered and oriented mouse within the arena. In addition, scalar values related to the kinematics of the mouse, such as velocity, mouse width, mouse length, mouse height, mouse centroid, etc are computed and aligned to the extracted depth frames when this cell is run (Figs. 4e).

    e. Run the Validation Tests cell, which executes extraction validation tests. The validation tests check the extracted sessions for mice that are unusually still, which can occasionally confuse the modeling downstream (which expects some degree of behavioral dynamics), mice missing from the images, size anomalies, dropped frames, and scalar anomalies. These flags are described in detail in the GitHub wiki.

CRITICAL: Warning flags indicate the sessions need additional examination and the session may need to be re-extracted or excluded from the downstream modeling.

    f. Review the quality of processed, extracted videos by running the Review Extraction Output cell. This enables a further, qualitative visual assessment of the extracted and aligned mice to reveal whether they are noisy, subject to wall reflections, incorrectly flip-corrected, etc.

CRITICAL: Analyzing corrupt images or incorrectly oriented mice will corrupt the downstream modeling. Users should remove such sessions from further analysis.

    g. Aggregate all the extracted results into one folder by running the Aggregate the Extraction Results cell. After running this cell a folder called aggregate_results and a moseq2-index.yaml file are generated and placed in the project folder. The aggregate_results folder holds all the data needed for the rest of the analysis pipeline and the moseq2-index.yaml contains metadata and file paths for the extracted files.

CRITICAL: when aggregating data in this step, the moseq2-index.yaml file is re-generated and will overwrite the existing file. If users wish to keep the original file, it should be renamed and moved before running this cell.

    h. Label each session with its experimental class by running the Assign Groups cell (Fig. 5a). Each session is a contiguously recorded set of images associated with a particular mouse in a particular experimental condition; a given experiment typically has many such sessions that comprise the complete experiment (often corresponding to many mice). This tool allows users to "name" each session to keep track of which mice were run under what experimental conditions. Sessions can be sorted for ease of labeling; click on a column name to sort the table by values in the column, and click the filter button to filter the values in a column (Fig. 5b). Click on the session to select the session. To select multiple sessions, click the sessions while holding the CTRL/COMMAND key, or click the first and last entry while holding the SHIFT key. Enter the group name in the text field (Fig. 5c, label 1) and click the Set Group Name button (Fig. 5c, label 2) to update the group column in the table for the selected sessions (Fig. 5c, label 4 and label 5). Click the Update Index File button (Fig. 5c, label 3) to save current group assignments to moseq2-index.yaml. The interactive widget in this cell writes the assigned labels to the moseq2-index.yaml.

CRITICAL: Running this cell requires that all the sessions have a metadata.json file containing a session name; this .json file is generated by the MoSeq2 acquisition tool described above, and as such MoSeq2 is only capable of modeling depth data that is acquired through the associated acquisition code.

    i. Check for outliers in scalar values by running the Further Extraction Diagnostics cell. The mean and standard deviation of the selected scalar values (i.e., mouse size, velocity) are plotted in adjacent columns (e.g., Fig. 4f). Aberrations in scalar values can often indicate pathology in the underlying data, or mice whose behavior is out of normal bounds (i.e., completely passed out). If you wish to identify which session is associated with an aberrant value, hover your cursor over the specific data point in question; this will return the actual values associated with that point, and the unique identifier (UUID) associated with the session responsible for that data point.

    j. (Optional). If you wish to remove a session based upon the UUID, specify the full or partial UUID in the target_uuid variable field and run the UUID lookup cell, which will allow you to look up the metadata (i.e., filename, file path) associated with the session you want to eliminate/edit.

CRITICAL: Mouse sizes i.e., length, width, and height across sessions should be similar. Outlier sessions where mouse size is clearly different from the group should be excluded from further analysis. Failure to do this will result in "syllable splitting," in which

<span style="color:red">the same behavior is assigned to two different syllables based upon differences in size. MoSeq is typically tolerant of size differences in the 15-20% range, which corresponds to the typical size variation associated with mice of the same approximate age.</span>

4. Model the extracted depth data. Timing ~variable

    a. Dimensionally reduce the extracted depth data by running the Fitting PCA cell (Fig. 10a). This step learns a low-dimensional representation of mouse pose to increase computational efficiency and reduce noise present for the modeling step. Table 4.2 shows the key parameters in the PCA step.

    b. To visualize the principal components (PCs), run the Visualize PCA results cell. The top 10 PCs should explain around 90% of the variance in the dataset (Fig. 6b) and should look smooth and well-defined (Figs. 6b and 6c).

<span style="color:red">CRITICAL: The 10 PCs = 90% variance explained rubric depends upon having enough data to effectively model (i.e., at least ~ 1 million total frames). If the PCA step looks over smoothed (meaning fewer than 10 PCs explain 90% variance), users can adjust the gaussfilter_space parameter to decrease the amount of smoothing applied to the images at this step.</span>

    c. To compute the PC scores for each frame, run the Computing Principal Component Scores (Fig. 10b). This cell produces a 10-dimensional time series of PC scores for each session; the processing time for training and applying PCA generally increases with recording duration (Fig. 6d).

    d. Compute the model-free changepoint duration histogram by running the Compute Model-free Changepoints cell (Fig. 10c). Changepoints describe moments in time when one action transitions into another, as determined by a model-free changepoints detection algorithm. These changepoints data guide selection of a parameter called "kappa" later in the protocol, which determines the global duration distribution of syllables. Typically, the distribution of changepoint durations in a mouse experiment is smooth, right-skewed, and peaks around 0.3 seconds (Fig. 7b).

    e. Generate a preliminary set of fit AR-HMM model(s) by running the Set Model Parameter cell, the Train Model cell and then the Get Best Model Fit cell. Note that you will run these cells twice. In the first iteration (described in this step and step 4f), a "kappa scan" will be performed, which will enable users to identify the kappa value to be used for final modeling. In the second iteration (described in step 4g), this kappa value will be used to model the data.

    The Set Model Parameter cell defines a set of default parameters used by the model fitting procedure. In general (with the exception of the kappa and num_iters parameters) these parameters should not be altered, but experienced users can change relevant parameters in this cell.

In the first iteration, set the kappa parameter to "scan" and leave the num_iters parameter at its default of 100. Once a value of kappa has been determined (by looking at the output of the Get Best Model Fit cell), change the kappa to either "none" or to the desired integer kappa value from the output of the Get Best Model Fit cell, and increase the num_iters parameter to 1000 (see step 4g).

Parameters of interest (described in Table 4.3) include:

Use_checkpoint: this allows users to save intermediate analysis during the modeling step. It is set by default to False.

npcs: this indicates the number of principal components used by the model. Default is set at 10; note that 25 PCs are actually computed over the imaging data, which represent the maximum that can be submitted to modeling.

max_states: this indicates the maximum number of syllables used by the model to explain the data. Note that MoSeq uses Bayesian non-parametric techniques to learn the optimal number of syllables in any given dataset; this parameter places a cap on the number of possible learned syllables.

robust: each syllable is associated with its own noise distribution. By default (True) MoSeq will model each syllable using a student's t distribution, which tends to aggregate outlying behavioral instances into fewer syllables than the alternative noise model, a gaussian distribution (default = False). It takes longer to model behavioral using the student's t distribution than using the gaussian distribution.

separate_trans: this defines whether all the sessions in the experiment learn the same underlying transition matrix, or whether each class learns its own transition matrix. Note that this does not mean that all classes will have the same empirical transition matrix. Default is set to False.

CRITICAL: setting this parameter to True has the potential to introduce class leakage into downstream analyses. The amount of data being modeled is also effectively reduced so the model may not be sufficiently trained.

kappa: this value determines the duration of each syllable. The mean syllable durations increase as kappa value increases (Fig. 7c). By default kappa is set to "none," which means that kappa will be equal to the number of frames in the dataset. We have empirically observed that for most datasets a kappa value equal to the number of frames will yield a syllable duration distribution that is well matched to the mean of the changepoints distribution.

num_iter: this defines the number of training iterations (i.e., Gibb's sampling steps) the fitting procedure will run before terminating. This defaults to 100 for the initial scan. Users should change this to 1000 for the second run of this cell, once a kappa has been determined; this enables optimal "burn-in" of the final model (Fig. 7d and 7e).

select_groups: this parameter allows users to model a subset of classes in their project. Default is False; if set to True a prompt will appear at the train model step asking the user to specify which classes are to be modeled.

f. Visualize the output of the initial kappa scan by running the Get Best Model Fit cell. This cell will output a plot showing the syllable duration distributions for a wide variety of kappa values, as well as the model-free changepoints distribution. The cell also outputs the kappa value used by the model that best matches the changepoints distribution. This will allow a user to select a value for kappa when the model is re-run in its final configuration. The model syllable duration distributions should generally match the model-free changepoint duration distributions. The mean syllable duration from the target model should match that of the model-free changepoint duration. Based on prior work, the target median syllable durations should be 0.3-0.4 seconds.

CRITICAL: MoSeq does not — and is not expected to — return a syllable duration distribution that exactly matches the changepoints distribution. Users should select a kappa that best matches the mean of the changepoints distribution, which often results in a larger rightward tail in the syllable distribution relative to the changepoints distribution. These distributions will be visualized in the Get Best Model Fit cell at the end of this notebook.

g. Re-run the model in the final configuration. Set num_iter to 1000. Set base_model_path variable field to a new folder that keeps the final model(s). Set kappa in the Set Model Parameters cell to None or a specific kappa value. When config_data['kappa'] is set to None, kappa is set to the total number of frames during model training. In general, None should return appropriate results. Based on prior work, the optimal kappa value is nearly always within an order of magnitude of the total number of frames. When config_data['kappa'] is set to a specific integer, kappa is set to that value during model training. Rather than running a single model at this stage, typically we run ~100 models and compare their outputs based upon their log-likelihoods, selecting a model with a median log-likelihood to further explore. To do this, set the objective variable field to "median_loglikehood" in the Get Best Model Fit cell to find the model that has the median loglikehood among all the 100 trained final models.

**Procedure 3: Data visualization and analysis**

Table 2.2 and Table 2.3 show the project directory file structure after training one model and three models at the beginning of Procedure 3. Note that the analysis notebook can visualize multiple models in parallel; instructions below describe procedures for a single, optimized model.

The data can be analyzed and visualized using either the Jupyter Notebook or the CLI. The Jupyter Notebook provides additional interactive visualization functionalities that are not available in the CLI. This protocol only describes the Jupyter Notebook for analyzing and visualizing the data; a detailed description of the CLI alternative can be found in the GitHub wiki (https://github.com/dattalab/moseq2-app/wiki/Command-Line-Interface-for-Visualization).

1. Start the MoSeq2 Analysis Visualization Jupyter Notebook. Timing ~2 minutes

CRITICAL: Anaconda users run steps a -f; Docker users run step g-h.

   a. (Anaconda users only) Open Terminal on the analysis computer. Terminal is typically available in the applications directory.

   b. (Anaconda users only) Activate the MoSeq Conda environment by entering the following command, assuming the name of the environment is moseq2-app:
      >>> conda activate moseq2-app

   c. (Anaconda users only) Navigate to the project directory in Terminal. To do so, enter the following command (replace desktop/project with the path to the specific project directory):
      >>> cd desktop/project

   d. (Anaconda users only) Check the contents in the project directory by opening the directory in the finder, or by entering the following command in Terminal:
      >>> ls

   Folder contents should be organized as indicated in Table 2.2 and Table 2.3.

   e. (Anaconda users only) Launch the Jupyter server by entering the following command in Terminal:
      >>> jupyter notebook

   This will open up a webpage that indicates all of the contents in the project directory.

f. (Anaconda users only) To start the notebook, click "MoSeq2-Analysis-Visualization-Notebook.ipynb" to start using the notebook.
g. (Docker users only) Once the Docker container is running, click the "OPEN WITH BROWSER" icon to open a webpage to run the Jupyter notebook. Input "moseq" in the password text field to display all the content in the project directory.
h. (Docker users only) To start the notebook, click "MoSeq2-Analysis-Visualization-Notebook.ipynb" to start using the notebook.

2. Validate the project. Timing ~20 minutes

a. To load the progress.yaml file associated with the trained model, run the Load Progress cell. This cell will also verify that the necessary paths (such as base_dir, config.yaml, moseq2-index.yaml, aggregate_results, pca files and models) for the analysis are recorded in the progress.yaml file (Fig. 11a).

b. To create a model-specific folder for analytical outputs, run the Setup Directory Structure for Analyzing Model cell. A model-specific folder will be created and the model will be copied to the respective folder after running the cell. Analysis results will be stored in the respective model-specific folders. This step will update the progress.yaml file to point to the model-specific folder.

c. Optionally, if multiple models were run and saved in the folder specified in the base_model_path field, you can specify the specific you wish to analyze by running the Override Paths to Specific Model. The cell will update the progress variables and the progress.yaml file. Users can select a model for analysis by simply inputting the file name for the new model in the desired_model variable field.

3. Generate statistics for a MoSeq model. Timing ~20 minutes

a. To aggregate frame-by-frame MoSeq outputs into a single dataframe, run the run the Compute moseq_df cell. moseq_df is a vertically stacked dataframe of values related to model results. The dataframe includes scalar values (e.g., mouse angle, position, height, length, centroid, velocity) and syllable labels aligned to individual frames, timestamps, and session metadata, such as the mouse and session names (Table 5.1). Because this dataframe includes frame-by-frame annotation of syllable usage as well as frame-by-frame scalar values, it is suitable for generating e.g., transition matrices, positional plots, etc. This dataframe (which is sufficient to recapitulate all analyses below, and which users can use for offline analysis of MoSeq outputs independent of this visualization package) will be used by subsequent cells for data visualization. Note that syllable numbers are assigned based upon usage; in this convention, the most frequently used syllable in a given experiment is referred to as Syllable 0, the second most used syllable is Syllable 1, etc. The cell output in the notebook shows a preview of the top 5 rows in the dataframe.

b.  (Optional) To export the scalar dataframe to CSV for offline analysis, run the Export moseq_df cell. CSV files will be saved into the model-specific folder.

c.  To generate a dataframe that contains average scalar values and the frequency of usage associated with each syllable (i.e., the average velocity associated with Syllable 24), run the Compute stats_df cell. stats_df is a dataframe grouped by the sorted syllable labels, experimental classes, and UUIDs (Table 5.2). This dataframe will be used in subsequent cells to support computing syllable-specific metrics. The cell output in the notebook shows a preview of the top 5 rows in this dataframe.

d.  (Optional) To export the scalar dataframe to CSV for offline analysis, run the Export stats_df cell. CSV files will be saved into the model-specific folder.

e.  To generate a MoSeq fingerprint plot (which includes a graphical description of scalar values as well as syllable usages for each mouse), run the Generate the MoSeq Fingerprint cell. MoSeq fingerprint plots (Fig. 11d) are a set of heatmaps that plot the distributions of kinematic parameters (e.g., position, 2D velocity, height, and length) and syllable usage using the moseq_df and stats_df dataframes. The heatmaps are intended to show a summary of behavior across sessions and experimental groups, thereby enabling quick quality control and identification of obvious differences between groups.

f.  To generate crowd movies and to label individual syllables based upon the movements observed during syllable expression, run the Interactive Crowd Movie Generation and Syllable Labelling Tool cell. For each syllable, a "crowd" movie (Fig. 8a, label 1) will be generated, allowing the user to observe the 3D behaviors specifically associated with expression of a particular MoSeq-identified syllable. A folder called crowd_movies that contains all the crowd movies (saved in .mp4 video files) and a syll_info.yaml file that holds syllable information are generated and placed in the model-specific folder. In addition, this cell enables users to label syllables based upon the behaviors observed in each crowd movie. Functionality in the cell is provided to manipulate the playback speed to make it easier to identify and distinguish behaviors (Fig. 8a label 2), as well as for users to specify a name and description for each syllable (Fig. 8a, label 3 and label 4). Click the Save Setting button (Fig. 8c label 5) to save the syllable name and description to syll_info.yaml file. Click Next button (Fig. 8c label 6) to save the syllable name and description and to go to the next syllable crowd movie. Click Prev button (Fig. 8c label 7) to go to the previous syllable crowd movie.

    CRITICAL: You must save the labels for each syllable by pressing the Save Setting button or the Next button; these labels are not automatically saved when switching from one syllable to another syllable from the syllable dropdown menu.

g.  Run the Generate Dendrogram cell to generate a hierarchically-sorted dendrogram (Fig. 11e), depicting the kinematic similarities amongst the observed syllables. The distance metrics that are used to identify kinematic similarities are based upon the auto-regressive matrices computed as part of the MoSeq modeling step.

h.  To generate usage plots (which depict how often each syllable is expressed by a given mouse or experimental group), Run the Interactive Syllable Statistics Graphing cell (Fig. 8b). Hovering over each point in this plot (which corresponds to a particular syllable) will reveal syllable-specific scalar and usage statistics, as well as the crowd movie associated with the syllable.

    Under the "grouping" drop-down menu, you can select whether to plot syllable usages based upon the experimental class to which the mice belong, the specific session, or the specific mouse. These usage plots can be exported (in svg format) by clicking the save icon to the right of the plot. The raw data that is used to generate this plot is found in the stats_df dataframe generated in step 3d; users who wish to generate custom representations of usage data can extract it from that dataframe. Drop down menus are provided that enable flexible sorting, different types of error bars, etc.

i.  (Optional) To generate plots depicting syllable duration, per-syllable velocity, per-syllable height, and average per-syllable distance to arena center, use the drop down Stat to Plot menu.

j.  To compute and visualize transition matrices depicting the empirically observed likelihood that any given syllable transitions into any other syllable during a given experiment, run the Compute Syllable Transition Matrices cell. This cell will output a transition matrix for each experimental class (Fig. 11f). The rows of each transition matrix represent an incoming syllable, while the columns represent the outgoing syllable. The value at a specific row and column position represents the frequency the incoming syllable transitions into the outgoing syllable (equivalent to the frequency of each bigram). This image is saved automatically to the Plots folder within each model folder.

k.  (Optional) To export the transition matrix to CSV format, run the Export Transition Matrices and Usages cell.

l.  To generate graphical representations of each transition matrix (i.e., state maps), run the Interactive Syllable Transition Graph Tool cell. This cell enables the user to plot syllable transitions within each group, and to also plot transition differences between two groups (e.g. up-regulation and down-regulation in syllable usages and syllable transitions). Hover over the nodes to display the associated syllable crowd movies and information about syllable-associated statistics. The edges linked to each node are colored by edge types (e.g., incoming, outgoing and bidirectional), and are displayed when the cursor hovers over the node. Plotting options are available in the dropdown menus (Fig. 8c, label 1-3). To save these plots in pdf format, click the "Save" icon at the top of the plot; plots will be saved to the Plots folder within each model folder.

Table 6 summarizes the commands used throughout Procedure 2 and Procedure 3.

**Troubleshooting**

The MoSeq2 package suite includes exception handling which indicate commonly occuring errors. Additional information regarding errors generated by the package, and approaches for dealing with common errors, are described in the MoSeq GitHub wiki (https://github.com/dattalab/moseq2-app/wiki). Table 7.1 describes common issues encountered during data acquisition; Table 7.2 and Table 7.3 describes common issues encountered during to use of the MoSeq2 package suite.

We have set up GitHub Issues section (https://github.com/dattalab/moseq2-app/issues) in the moseq2-app repository for users to search for existing solutions and report problems. The MoSeq website also provides access to a Slack channel, workshops and Q&A sessions to help users set up the MoSeq framework, troubleshoot and discuss project-related issues.

**Timing**

The acquisition software installation times indicated above are for a SuperLogics computer running Windows10 operation system (NVIDIA GeForce 1030 GPU, 3.7 GHz i3-6100 CPU, 8GB of RAM). The MoSeq2 package suite installation times are for a MacBook Pro (2.3 GHz 8-Core Intel Core i9 CPU, 32GB of RAM), running macOS Monterey 12.1 using Anaconda, and times in the pipeline are for that same computer using the full test dataset with 48 sessions. The expected timing includes typing the commands, going through extraction parameters for all 48 sessions, extracting sessions, running PCA and modeling.

Setting up the depth video recording apparatus will vary depending on the familiarity of assembling cages using 80/20 hardware. The total time needed to build the recording apparatus (i.e. the recording cage, the camera and the bucket) is approximately two to three hours; after assembly the bucket needs to air dry for 24 hours after painting. The time needed to set up the acquisition software is approximately two to three hours.

The timing needed to install the MoSeq2 package suite depends on the environment and the installation route. MoSeq2 package suite can be installed using MoSeq2 Docker image or Anaconda. The total time needed to install the MoSeq2 package suite varies depending on the chosen installation route. In general, the MoSeq2 package suite can be installed in 30 minutes to one hour.

The time needed to define the session-specific parameters for extracting the depth videos depends on the quality of the depth videos, as well as the variation between sessions if multiple recording apparatuses are used in the experiments. Examining the extracted videos and the raw data may be necessary if the extracted results have major issues, such as a large amount of noise and reflection, too many dropped frames in the recording, and/or mice being mostly still. It should take approximately 20 mins to set up the notebook for extraction and modeling. Finding parameters and extracting the first 20-minute video takes approximately 15 to 20 minutes, with subsequent videos (whose parameters may be similar/identical to the first) taking less time.

The time needed to label each experimental group in the group assignment step depends on the total number of groups and the chosen naming convention for the sessions in the metadata. The time needed to extract the depth videos, train PCA, apply PCA and train the AR-HMM models depends on the total duration of the depth recording and the specifications of the analysis computer. The process can be made faster when using parallel computing on HPCs such as Slurm; detailed documentation for parallel computing and MoSeq is on the GitHub wiki page. Time needed for dimensionality reduction and modeling step varies depending on the amount of data and the number of iterations for the modeling step and the type of model. Fig. 6d and 7f benchmark the approximate processing time.

**Anticipated results**

As part of the MoSeq2 package we provide an example behavioral dataset that enables exploration of the method and its outputs. There are 48 sessions in this dataset, corresponding to 48 different mice; in half of the sessions, mice were injected with amphetamine (2 mg/kg) while in the other half, the mice were injected with saline as a control. Below, we describe the expected outcomes when using these reference data, including the files that are expected to be generated and their contents.

Procedure 1: Data acquisition

After a depth recording has ended (Step 3c in Procedure 1), the acquisition software produces a folder named with the current date and time in the format of "session_<datetime>". The folder contains the raw depth video in a depth.dat file, timestamps for each video frame in a depth_ts.txt file, and optionally an rgb.mp4 video file if the color stream is also saved. These files are already provided within the example dataset, thereby enabling new users to explore MoSeq and examine its outputs without having to acquire initial depth data.

Procedure 2: MoSeq extraction and modeling

After the 3D depth video of mouse behavior is collected and organized into experiment-specific projects (as provided in the example dataset), the Extraction and Modeling Notebook generates a progress.yaml file and sets up the progress variables (Fig. 9b) to track the analysis pipeline progress (Step 2b in Procedure 2), generates a config.yaml file to store the configurable parameters (Step 2c in Procedure 2) and downloads a flip classifier for the extraction step (Step 2d in Procedure 2). At the beginning of the extraction steps, the Arena Detection Tool (Step 3a, 3b and 3c in Procedure 2 finds optimal parameters to detect the floor of the arena, perform background subtraction, and extract the mouse. The Arena Detection Tool generates a session_config.yaml file to store session-specific configuration parameters. Once parameters are chosen and validated by the user, the mouse is then centered and oriented facing to the right, producing a results_00.h5 file (Step 3d in Procedure 2, Fig. 9c). This file also contains basic kinematic values describing the behavior of the mouse, such as velocity and height (Fig. 4e). Users can run extraction validation tests (Fig. 9d) and review the extraction results (Fig. 9e) to ensure the quality of the extracted videos. Aggregating the data (Step 3g in Procedure 2) copies and renames the results_00.h5 files from each recording in the experiment into a new folder called aggregate_results, which enables dimensionality reduction and modeling (Fig. 9f). The moseq2-index.yaml file is then produced to store session metadata and group labels, which are specified by the user in the Assign Group cell (Step 3h in Procedure 2).

To fit a MoSeq model to the data in a reasonable amount of time, and to reduce contamination with unnecessary sources of noise, the dimensionality of the centered and oriented mouse pose time series is reduced with PCA (Fig. 10a), producing a pca.h5 file that contains the top 25 principal components (Step 4a in Procedure 2) of extracted images. The depth images then are projected to the principal components (Fig. 10b), producing a pca_scores.h5 file that contains the top 10 PCs over time for each recording session included in the experiment (Step 4c in Procedure 2). To help set the kappa parameter, model-free changepoint durations are computed from PC scores (Fig. 10c), producing a new changepoints.h5 file (Step 4d in Procedure 2).

The final steps of Procedure 2 fit an AR-HMM to the PC scores of each recording session (Step 4e, 4f in Procedure 2), producing a pickle file (i.e., model.p) containing model-specific parameters and syllable labels for each session (Fig. 10d). The fit is validated by comparing the resulting syllable duration distribution to a model-free changepoint distribution (Step 4g in Procedure 2).

Procedure 3: Data visualization and analysis

Once a validated model is trained, summary DataFrames, moseq_df (Fig. 11b) and stats_df files (Fig. 11c) are generated for data analysis. A MoSeq fingerprint plot (Fig. 11d) is generated to summarize the key scalar values and syllable usages across sessions. moseq_df stores scalar values and syllable labels for each frame in the sessions and stats_df averages moseq_df by syllable and group (Step 3a, 3c in Procedure 3). The MoSeq fingerprint plot is a histogram that shows the distribution of the kinematic values and syllable usages (Step 3e in Procedure 3).

Once a validated model has been trained on the data, crowd movies for each syllable are generated so that a human observer can label them (Step 3f in Procedure 3), producing a folder containing a movie file for each syllable called crowd_movies and a syll_info.yaml file that stores the user-generated syllable names and descriptions. A dendrogram (Fig. 11e) is produced to show the kinematic similarities amongst the observed syllables (Step 3f in Procedure 3).

In an experiment containing more than one experimental group, it is often useful to compare the frequency with which each syllable is used (i.e., syllable usage) and other kinematics metrics (e.g., 2D velocity, duration, height etc.) to understand how experimental manipulations affect behavior. Syllable statistics are plotted for each experimental group in an interactive plot (Step 3h, 3i in Procedure 3). In this plot, statistical tests are performed for each syllable when the Sorting dropdown (Fig 8b label 3) is set to "Difference" to determine if two groups are significantly in the plotted syllable statistics. Transition matrices and bigrams (Fig. 11f and Fig.8c) are plotted to reveal how transition structure changes between experimental manipulations (Step 3j, 3l in Procedure 3).

**Data and code availability**

Data and code available via http://www.MoSeq4all.org.

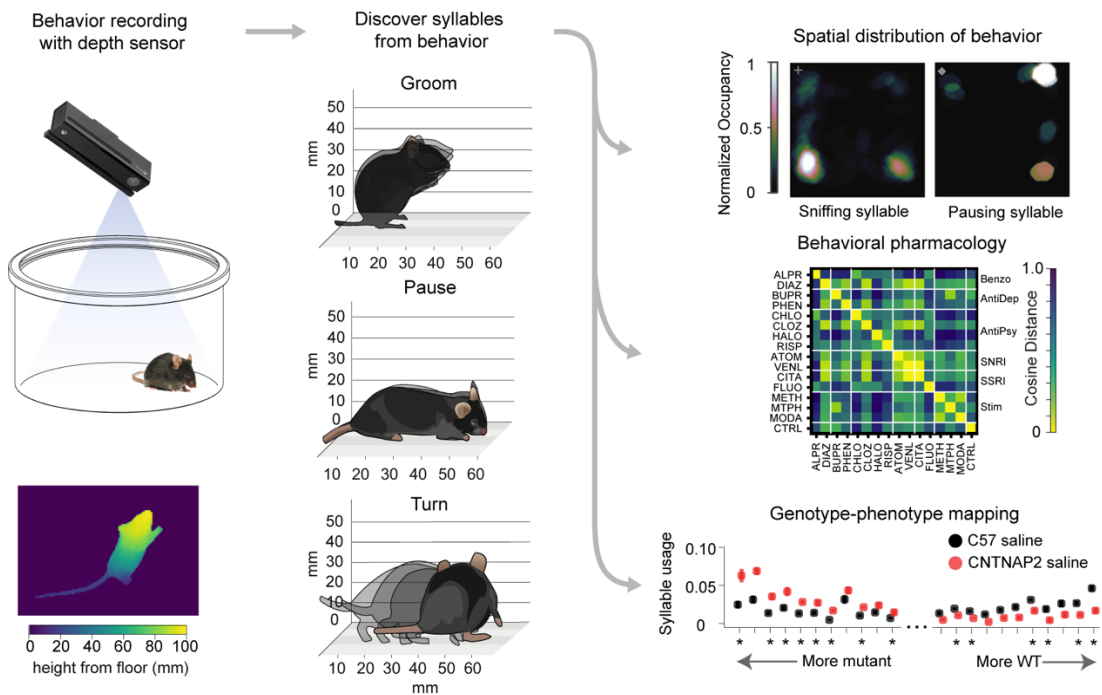

Figure 1. Overview of the MoSeq pipeline and typical applications.

Left: Schematic depicting a typical experimental setup. Top left: a Kinect2 depth camera is used to record depth videos of mice behaving in simple arenas like the depicted open field. Bottom left: a depth image of a mouse. The color of each pixel indicates the height relative to the floor, where brighter colors indicate greater heights. Middle: Example syllables (with human annotated labels) discovered by MoSeq. Right: example applications of the pipeline, demonstrating the ability of MoSeq to identify syllables and sequences that are changed as a consequence of environmental, pharmacological and genetic manipulations.

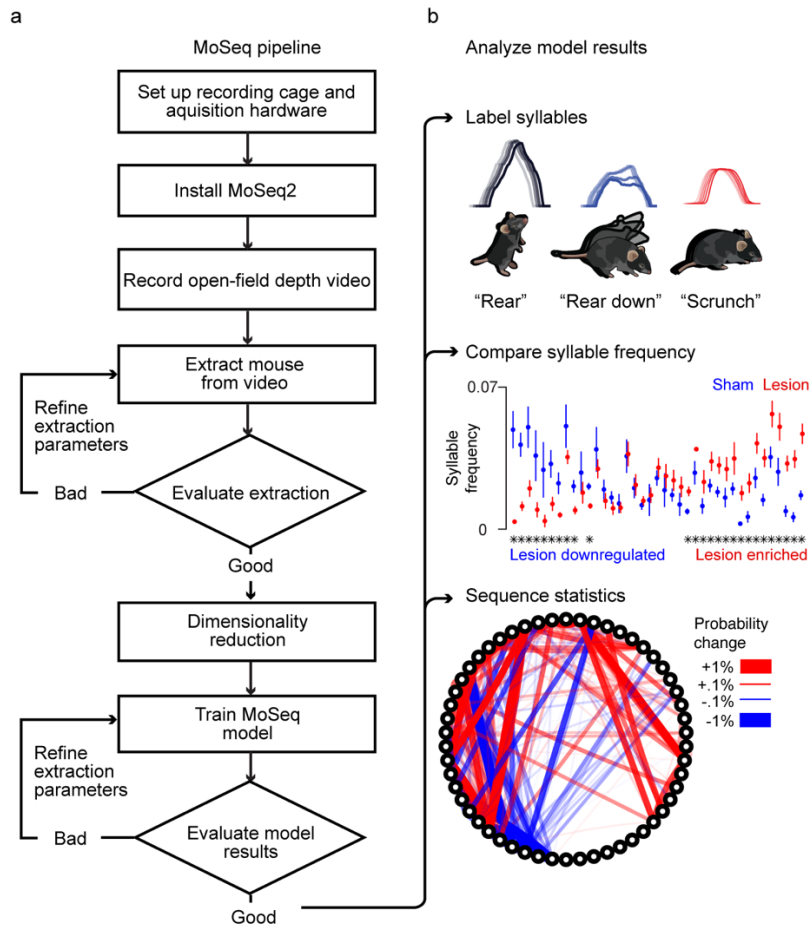

Figure 2. Schematic of the MoSeq pipeline.

a. Schematic of the data processing pipeline. Raw depth videos of behaving mice are recorded; the image of the mouse is then extracted (i.e., distinguished from the arena in which the mouse is behaving), and then oriented such that the mouse's nose always points to the right and the mouse's body is located in the center of the image. After the extraction and alignment steps, the mouse depth images are subject to dimensionality reduction and then submitted to the AR-HMM model, which learns a syllable label for each frame of depth data.

b. Several example outputs of the MoSeq pipeline. As illustrated from top to bottom, tools are provided that allow users to assign semantic meaning to syllables (by visualizing each syllable in movie form), to compare how often each syllable is used across experimental conditions (here with syllables arrayed on the X axis, usage of each syllable on the Y axis, and the two different conditions in blue and red), and to characterize the transition structure that governs how syllables are strung together over time.

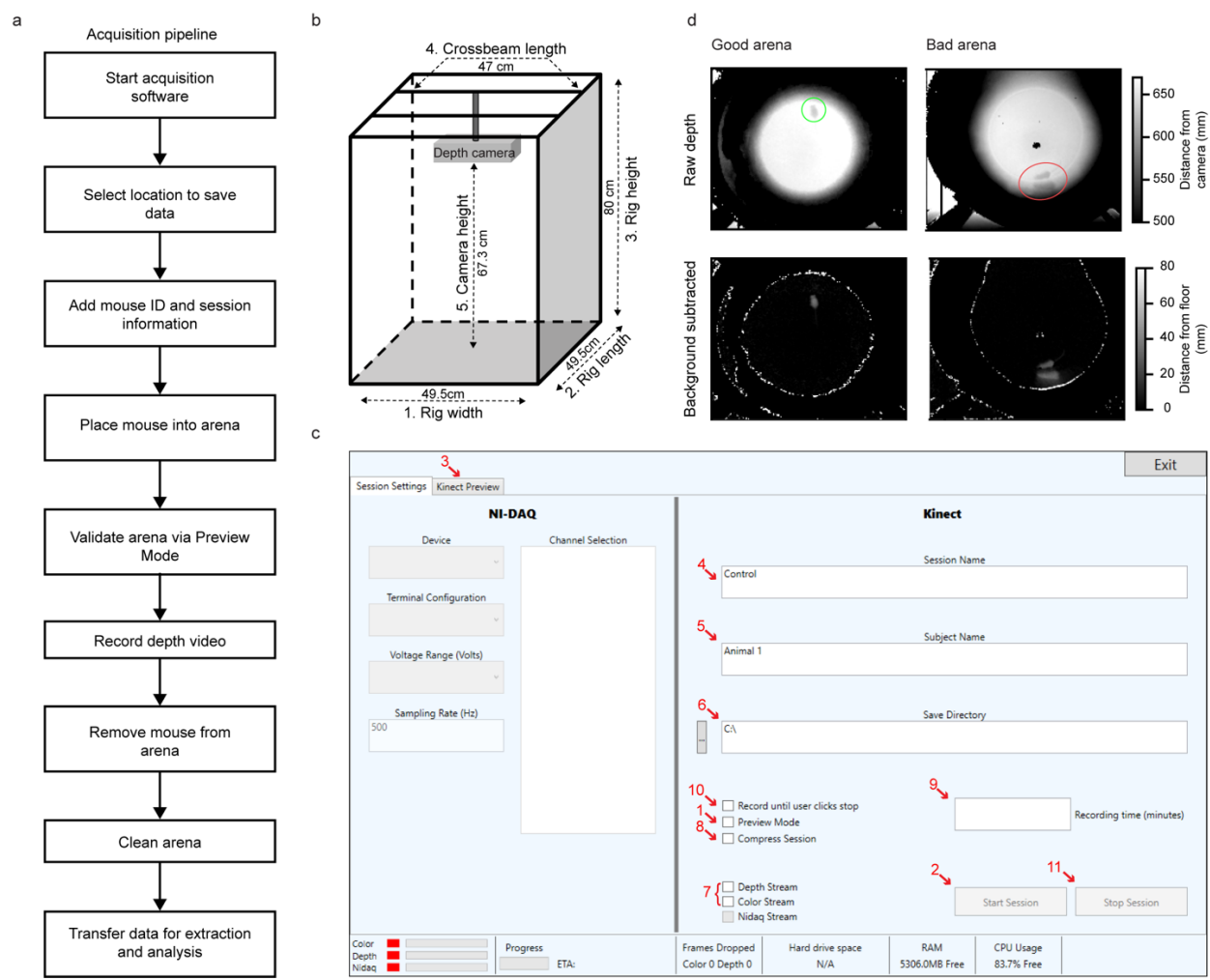

Figure 3. Data acquisition overview.

a. Schematic of the acquisition pipeline (corresponding to Procedure 1).
b. Diagram of the recording cage. A 49.5 cm wide (1), 49.5 cm deep (2), and 80 cm tall (3) cage suspends a Kinect2 depth camera 67.3 cm above the ground (4) via a 47 cm crossbeam (5). The cage is constructed using 80/20 aluminum beams.
c. Screenshot of the graphical user interface provided to users for performing data acquisition (corresponding to Procedure 1). Numbers refer to: 1) checkbox that switches the program into preview mode for testing and calibration; 2) button to start recording once a video source and recording duration have been selected; 3) tab that displays the depth and color video sources; 4) input text field for session or experiment type; 5) input text field for mouse ID; 6) file dialog for the data saving path; 7) checkboxes that enable the depth or color video sources; 8) checkbox to compress data at the end of the recording; 9) input text field for configuring the recording duration; 10) checkbox that allows timer-free recordings; 11) button to stop recording manually.
d. Depth images depicting the raw depth images (top) and the background subtracted image (bottom) from a properly prepared arena (left) and a poorly prepared arena (right). The green circle highlights the mouse in the properly prepared arena, while the red oval highlights the mouse in an arena that had not been sufficiently sanded. Notice that the image of the mouse is reflected in the walls of the poorly prepped arena.

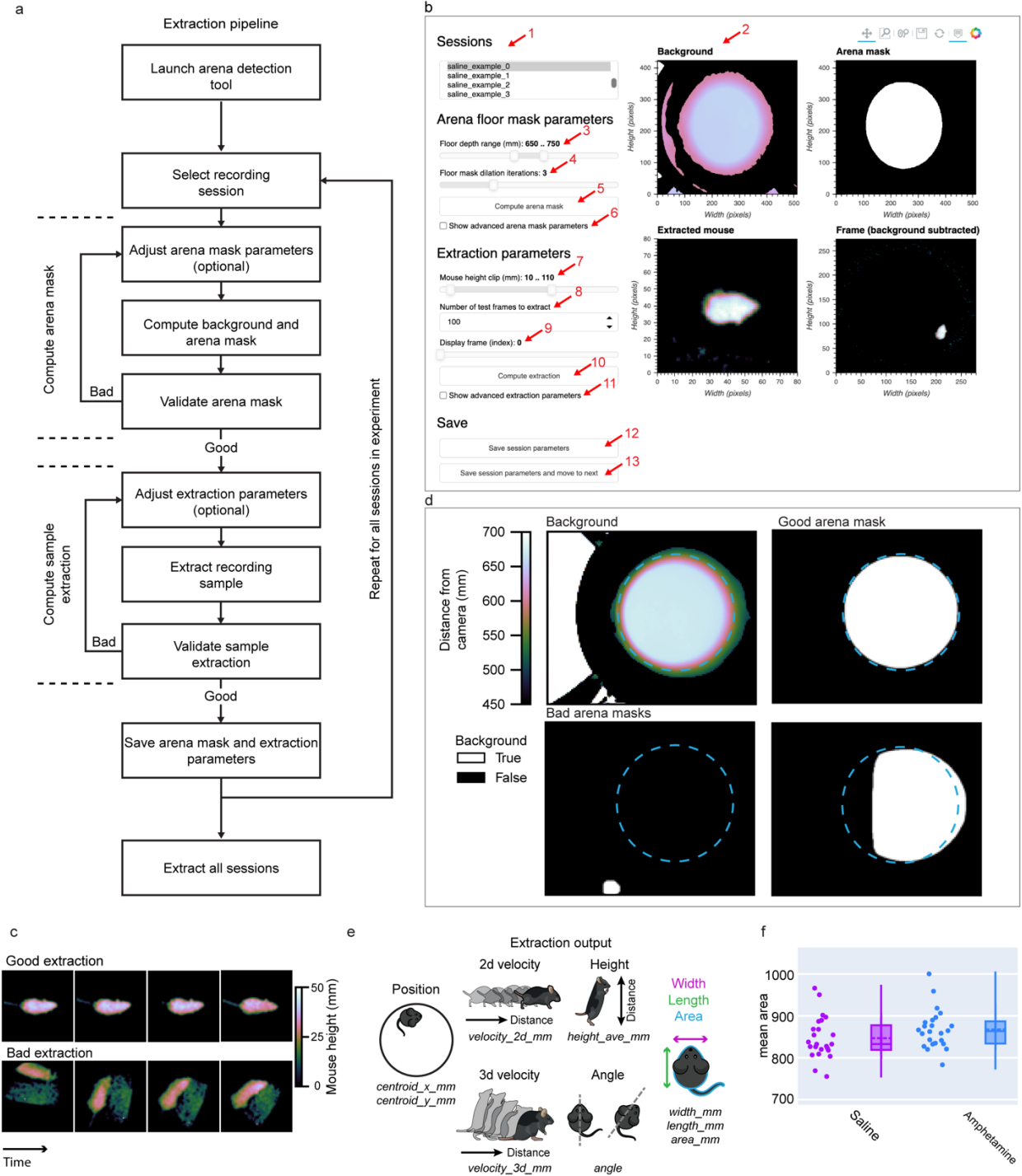

Figure 4. Overview of depth video processing and extractions steps.

a. Schematic of extraction pipeline (corresponding to Step 3a-3d in Procedure 2).
b. Screenshot of the Interactive arena detection tool. Numbers refer to: 1) the session selector, which indicates the current session being extracted; 2) the dynamic visualizer, which (given a set of parameters) visualizes the background, arena mask, extracted mouse and a background-subtracted preview frame; 3) a slider that specifies the floor depth range; 4) a slider that specifies the number of floor mask dilation iterations; 5) a button that computes the arena mask; 6) a checkbox that shows additional extraction parameters; 7) a slider that specifies the mouse height range; 8) a selector that enables users to specify the number of test frames to extract; 9) a slider that specifies the frame to display; 10) a button that, when pressed, initiates image extraction; 11) a checkbox that, when checked, reveals additional extraction parameters; 12) a button to save session parameters; 13) a button to save session parameters and move to the next session; note that this button automatically computes the arena mask for the next session.

c. Example frames after mouse extraction. Top: a series of depth frames consisting of centered and aligned mice from a properly-prepared arena. The mouse is centered and aligned rightward with minimal noise. Pixel values indicate height from the floor, with brighter colors indicating larger values. Frames from left to right indicate elapsed time. Bottom: a series of depth frames from an arena that had been poorly sanded. The mouse is neither centered nor aligned due to reflections and noise.

d. Example arena masks. Top left: the arena background (as shown in panel b). Brighter colors indicate distances that are farther from the camera. The dotted cyan circle indicates the floor of the arena. Top right: an example arena mask that is well fit to the floor of the arena. Bottom row: two examples of arena masks poorly fit to the floor of the arena.

e. Kinematic variables that are extracted by the MoSeq framework in the extraction pipeline. Italicized text indicates the variable names associated with each kinematic variable, as found in the output files of the extraction pipeline.

f. Example output after kinematic variables are computed post-extraction. In this example, swarmplots and boxplots were generated showing the distribution of average mouse sizes for two experimental groups, computed using the number of pixels occupied by the mouse.

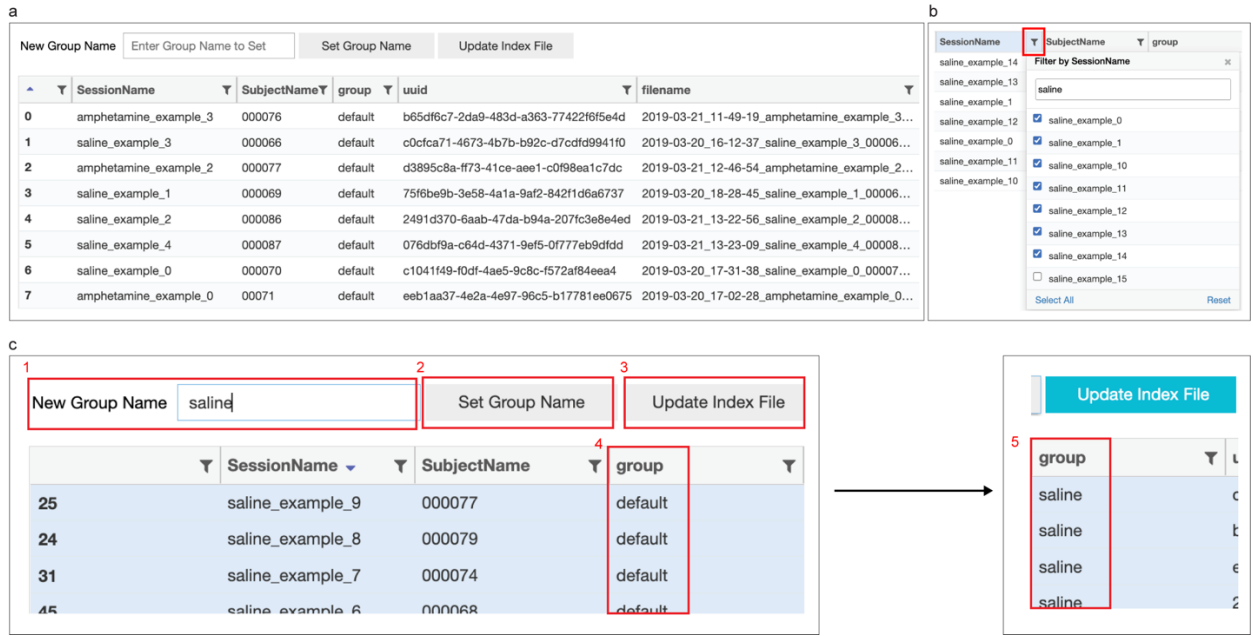

Figure 5. Overview of tools provided to assign imaging sessions to different experimental groups.

a. Screenshot of the experiment group assignment tool (corresponding to Step 3h in Procedure 2). Each row of the table displays the metadata for one recording session.
b. Screenshot of the filtering and selection functionality, demonstrating the selection of specific sessions by name, to assign a new group label.
c. Screenshot showing the process of changing an experimental group label. After typing a new group label into the text box, clicking the "Set Group Name" button (1) updates the group labels in the table (2, 3). The group labels are saved after clicking the "Update Index File" button (4) and updated in the table in the "group" column (5).

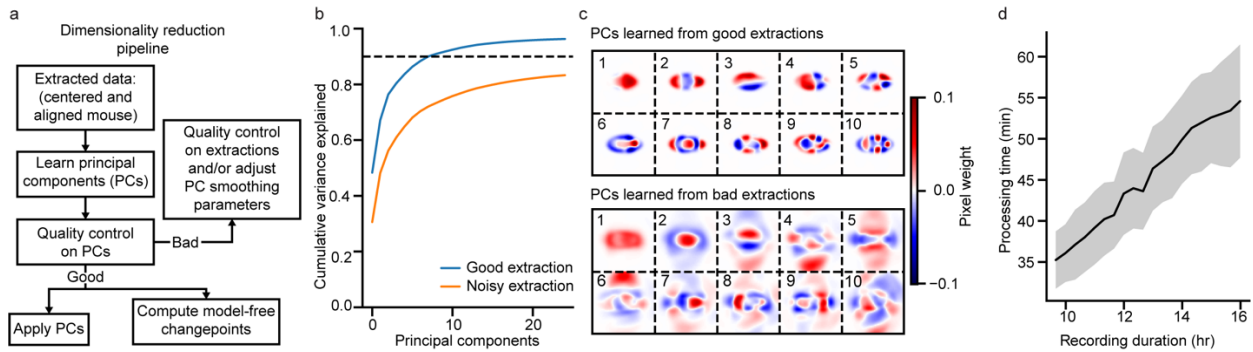

Figure 6. Overview of the dimensionality reduction pipeline.

a. Schematic of the dimensionality reduction pipeline (corresponding Step 4a-4d to Procedure 2).
b. Cumulative explained variance for an example good extraction (blue) and a noisy extraction (orange) for the top 25 principal components (PCs). The dotted black line indicates a 90% of the cumulative variance explained threshold, which in a good extraction should be explained by around 10 PCs.
c. Pixel-wise values for the top 10 PCs, illustrating which depth pixels contribute to each PC. The top plots represent PCs computed using data collected from a properly sanded and painted black bucket, while those on the bottom represent PCs obtained from mouse imaging in a lightly sanded and unpainted bucket with significant reflections. Note that in the bad extraction examples there are significant PC weights for pixels that surround (but are not part of) the mouse.
d. Line plot showing the amount of time needed to run the dimensionality reduction step with varying amounts of data. Eight CPU cores and 20GB of memory were used to generate this plot. Error bars indicate standard deviation across multiple runs.

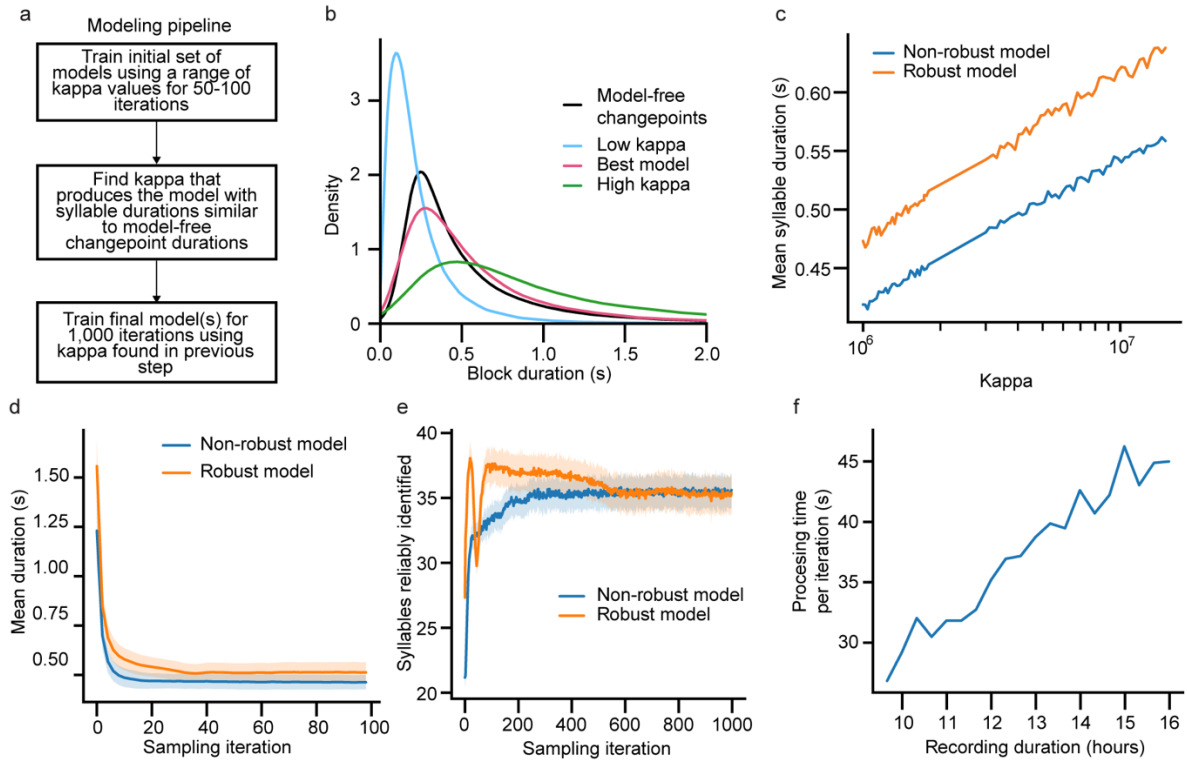

Figure 7. Overview of the MoSeq modeling pipeline.

a. Schematic of procedure used to pick an appropriate kappa and to generate an appropriate final MoSeq model for a given dataset (corresponding to Step 4e-4g Procedure 2).
b. Distribution of block durations for model-free changepoints (black) and syllable durations for models trained with a low kappa (blue), high kappa (green), or a kappa that best matches the mean duration of the model-free changepoints (orange).
c. Mean syllable duration for robust (orange) and non-robust (blue) models trained with different kappa values. The robust and non-robust models differ in terms of the noise distributions used to assign frames to syllables, with the robust model (which uses a student's t-distribution) being more generous in terms of agglomerating syllables relative to the non-robust model (which uses a gaussian distribution), which tends to generate more syllables for a given dataset.
d. Relationship between mean syllable duration and the number of Gibb's samples for robust (orange) and non-robust (blue) models. Error bars indicate standard deviation across recording sessions.
e. Relationship between the number of syllables used greater than 1% of the time and sampling iteration for robust (orange) and non-robust (blue) models. Note that this value stabilizes after about 500 iterations for a dataset with 16 hours of recording. Error bars indicate 95% bootstrap CI across recording sessions.
f. Relationship between recording duration and the processing time required to complete a single iteration of Gibb's sampling using the robust model. 8 CPU cores and 30GB of memory were used for this computation.

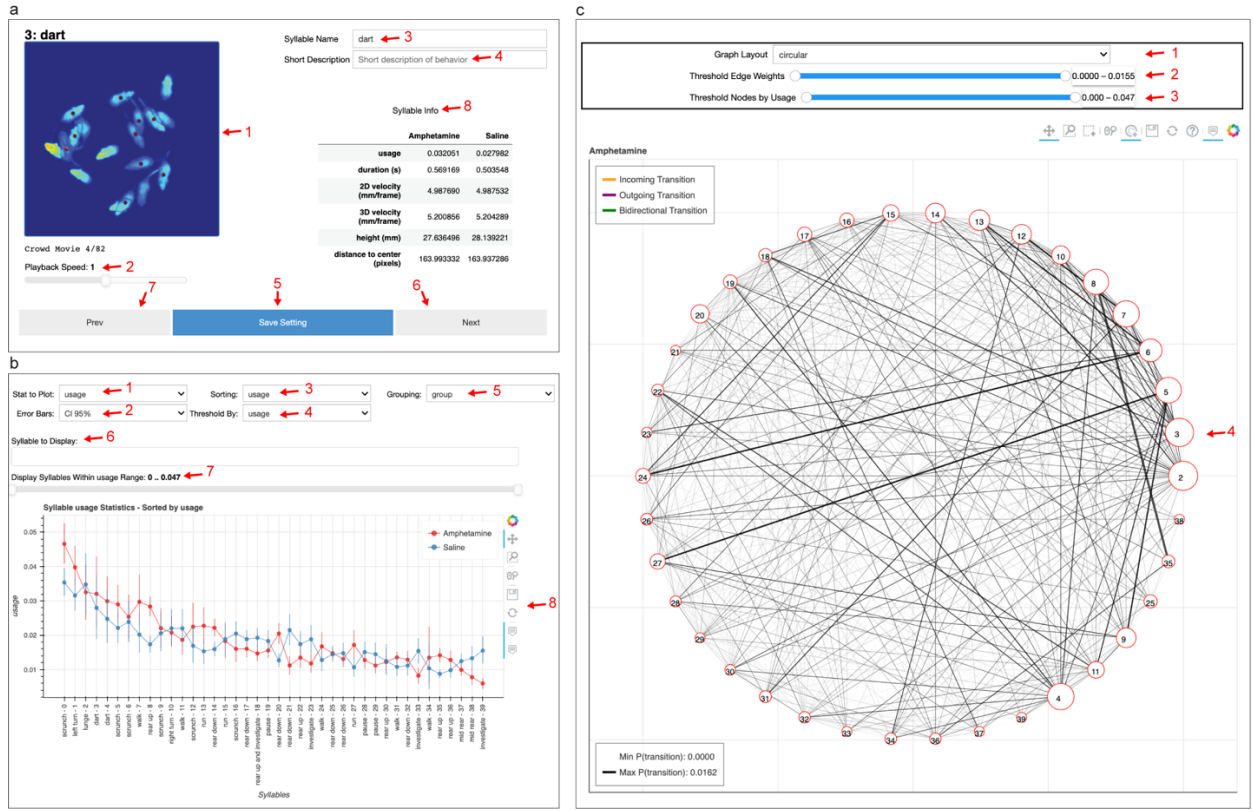

Figure 8. Overview of MoSeq analysis and visualization widgets.

a. Screenshot of the syllable labeling tool that enables users to visualize syllable crowd movies; this tool also provides a graphical interface that allows users, after inspection, to label syllables (corresponding to Step 3f in Procedure 3). Numbers refer to: 1) The visualizer that generates syllable crowd movies; such movies superimpose multiple examples of mice performing the selected syllable, enabling users to assign a natural language label to each MoSeq-identified unit of behavior; 2) the slider that enables users to specify the playback speed for the crowd movie; 3) the text field in which users can enter a descriptive label for each syllable based upon inspection of the crowd movies; 4) a similar input text field in which users can enter a short qualitative description of each syllable; 5) a button that saves the specified syllable name and the short description; 6) a button that allows users to annotate the next syllable; this button also saves the existing syllable name and short description; 7) a button that allows users to annotate the previous syllable; 8) a summary DataFrame that summarizes the kinematic statistics associated with the selected syllable.
b. Screenshot of the interactive syllable statistics visualization tool that enables users to compute and visualize different statistics related to syllable usage (corresponding to Step 3h in Procedure 3). Numbers refer to: 1) a dropdown menu specifying which specific syllable statistics to plot; 2) A dropdown menu that allows users to specify the type of error bar to be associated with each statistical analysis; 3) a dropdown menu allowing the user to select the specific sorting of the x-axis; 4) a dropdown menu that allows users to specify the type of threshold used to filter the statistics; 5) a dropdown menu allowing users to select how data are grouped; 6) an input text field enabling users to specify which syllable to display; 7) a slider allowing users to specify the specific threshold to be applied to a given type of data (as specified in label 4);  8) a plot that visualizes the user-specified syllable statistics. In this example, the reference dataset is composed of two groups of mice, one injected with saline and the other with amphetamine, and the use of each syllable is plotted. When the user hovers the cursor over any given node in this plot, the associated syllable crowd movies are displayed; this allows users to easily understand which behaviors are up- or down-regulated in a given experiment.
c. Screenshot of the Interactive syllable transition visualization tool that plots the transitions between syllables within each experimental group and enables users to also visualize differences in these transition statistics across experimental conditions (corresponding to Step 3l in Procedure 3).  Numbers refer to: 1) the dropdown menu allowing users to specify the layout of each transition graph; 2) a slider that allows users to specify the threshold for edge weights. Edges that lie outside the indicated threshold (and their associated nodes) are not displayed in the graph; 3) a slider that allows users to specify the threshold for how often each syllable must be used to be included in the graph; 4) the transition matrix plotted as specified by label 1.

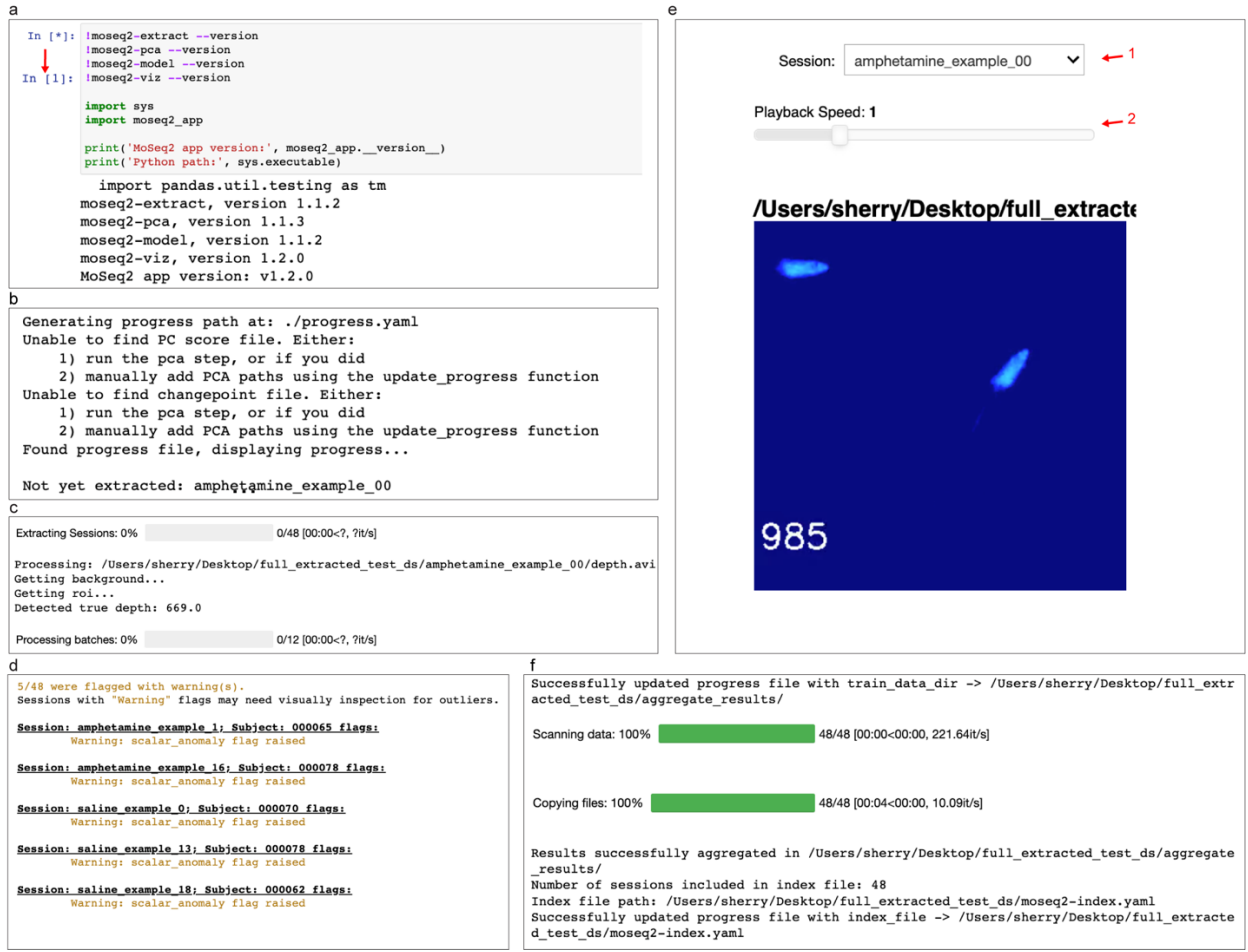

Figure 9. MoSeq extraction and modeling notebook outputs.

a. Screenshot of the Jupyter notebook output of the "Check MoSeq Versions" cell in procedure 1 that returns the version numbers of the MoSeq code being used. "In [*]" on the left indicates the cell is running; when this is converted (as shown with the red arrow) to "In [1]" the cell is done running. This will hold true for all cells in the notebook (corresponding to Step 2a in Procedure 2).
b. Screenshot of the Jupyter notebook cell "Set up or Restore Progress Variables" used to load the progress.yaml file (corresponding to Step 2a in Procedure 2).
c. Screenshot of the "Extract Session(s)" cell (corresponding to Step 3d in Procedure 2).
d. Screenshot of the "Run Extraction Validation Tests" cell (corresponding to Step 3e in Procedure 2).
e. Screenshot of the "Review Extraction Output" cell (corresponding to Step 3f in Procedure 2). Numbers indicate: 1) a dropdown menu that allows users to select the session to preview; 2) a slider that allows users to adjust the playback speed for the extracted video.
f. Screenshot of the "Aggregate the Extraction Results" cell (corresponding to Step 3g in Procedure 2).

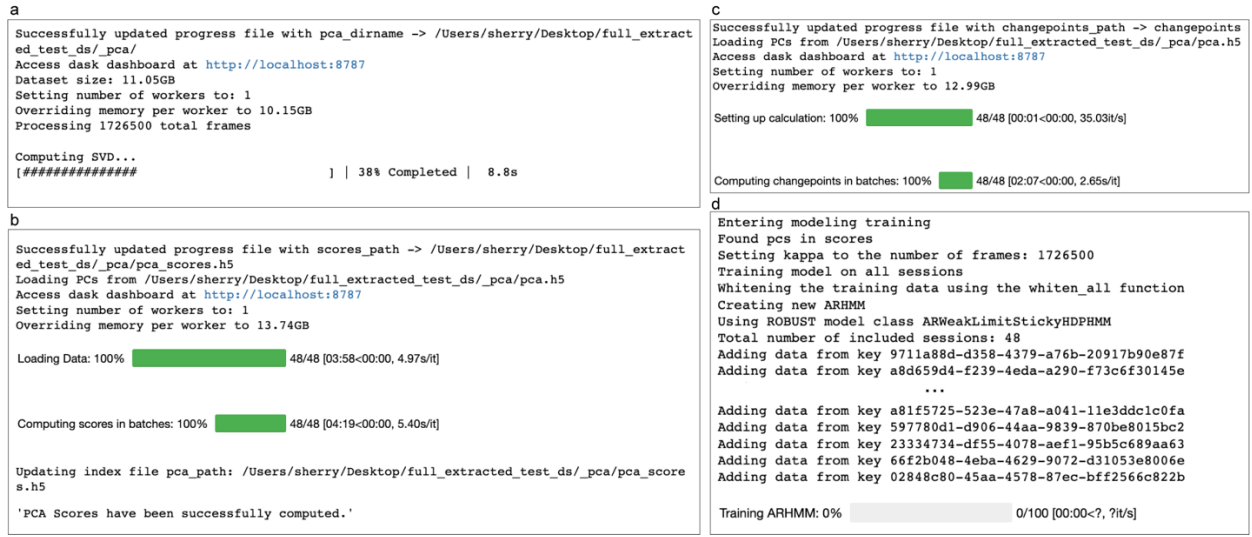

Figure 10. MoSeq dimensionality reduction and modeling step Jupyter notebook outputs

a. Screenshot of train PCA output of the "Fitting PCA" cell (corresponding to Step 4a in Procedure 2).
b. Screenshot of apply PCA output of the "Computing Principal Component Scores" cell (corresponding to Step 4c in Procedure 2).
c. Screenshot of compute model-free changepoint output of the "Compute Model-free Changepoints" cell (corresponding to Step 4d in Procedure 2).
d. Screenshot of the train model output of the "Train Model" cell (corresponding to Step 4g in Procedure 2).

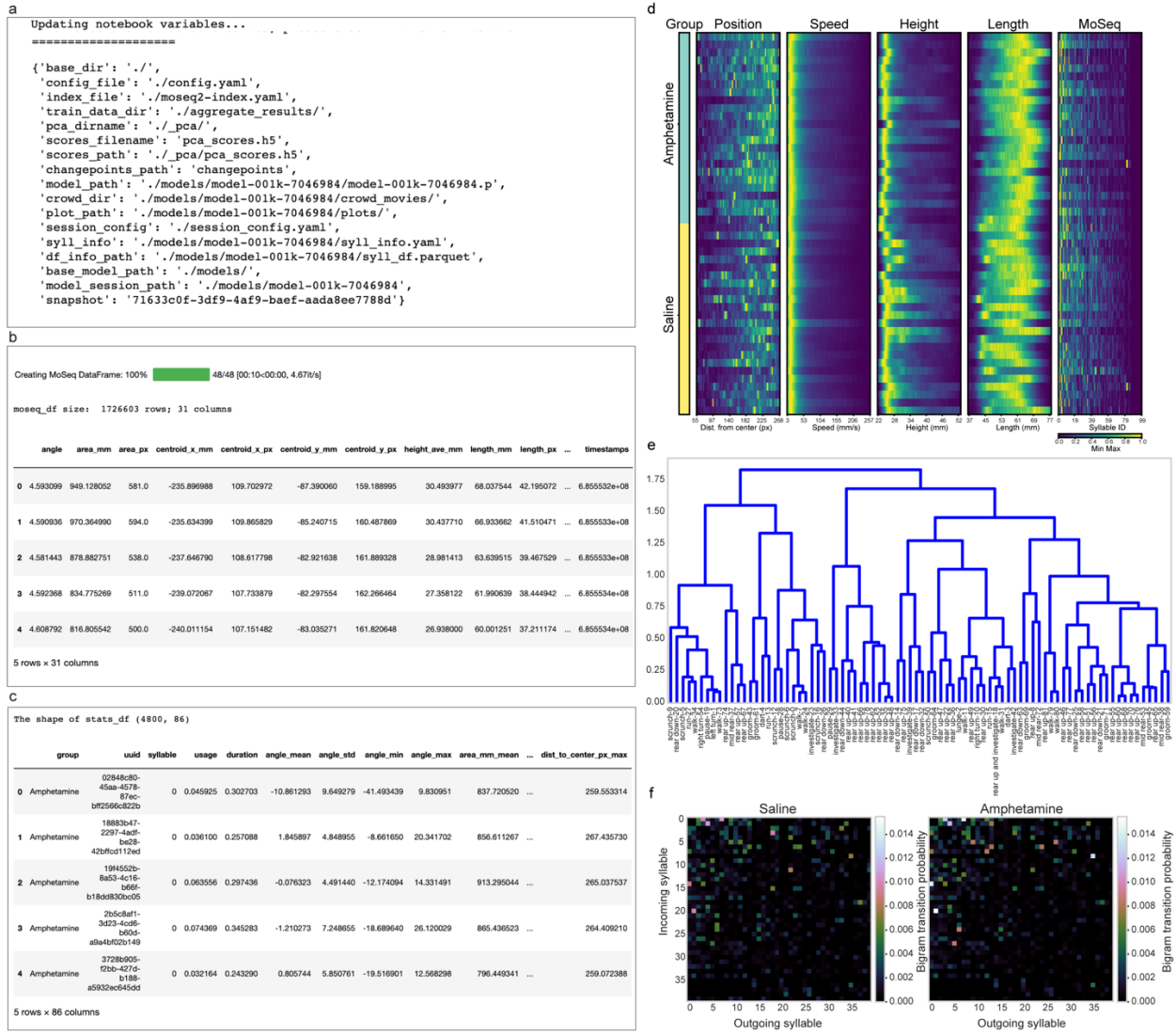

Figure 11. MoSeq analysis and visualization notebook outputs

a. Screenshot of the "Load Progress" cell at the beginning of the analysis and visualization notebook (corresponding to Step 2a in Procedure 3).
b. Screenshot of the "Compute moseq_df" cell (corresponding to Step 3a in Procedure 3).
c. Screenshot of the "Compute stats_df" cell (corresponding to Step 3c in Procedure 3).
d. A typical MoSeq fingerprint generated in the "Generate Behavioral Summary (Fingerprints)" cell (corresponding to Step 3e in Procedure 3).
e. A typical syllable similarity dendrogram generated in the "Generate Syllable Dendrogram" cell (corresponding to Step 3f in Procedure 3).
f. A typical syllable transition matrix generated in the "Compute Syllable Transition Matrices" cell (corresponding to Step 3j in Procedure 3).

# Table 1 | Bill of Materials

| Product | Part No. | URL | Amount | Unit Price | Description |
|---|---|---|---|---|---|
| Aluminum t-slotted beam | 1515 | https://8020.net/1515.html | 6 | $12.29 | 19.5" length, 15 series |
| Aluminum t-slotted beam | 1515 | https://8020.net/1515.html | 2 | $11.76 | 18.5" length, 15 series |
| Aluminum t-slotted beam | 1515 | https://8020.net/1515.html | 4 | $18.65 | 31.5" length, 15 series |
| Inside corner bracket with dual support | 14098 | https://8020.net/14098.html | 12 | $6.75 | 15 series |
| Bolt assembly | 3455 | https://8020.net/3455.html | 60 | $0.50 | 15 series |
| 90 degree inside corner connector | 3368 | https://8020.net/3368.html | 8 | $6.30 | 15 series |
| 10 Series 1" Narrow Horizontal Stanchion Base | 5860 | https://8020.net/5860.html | 3 | $31.45 | 10 series, for mounting camera and optogenetics fiber |
| Aluminum Tube Profile 1.000" Diameter | 5035 | https://8020.net/5035.html | 1 | $14.95 | 10" length |
| Bolt Assembly | 3471 | https://8020.net/3471.html | 10 | $0.73 | 1/4-20 x .875" Black SHCS with Washer and Slide-In Economy T-Nut - Centered Thread - Black Zinc |
| 1/4-20 Standard slide in T-nut | 3202 | https://8020.net/3202.html | 20 | $0.79 | 15 series, good for mounting |
| 1/4-20 Socket head cap screw | 3064 | https://8020.net/3064.html | 20 | $0.21 | 15 series, for use with 3202 |
| Ball Driver and "L" Wrench Set - Fractional | 6111 | https://8020.net/6111.html | 1 | $20.50 | for securing 14098, 3368, 3064, 3455 |
| Shallow tank | 14317 | https://www.usplastic.com/catalog/item.aspx?itemid=120721 | 2 | $107.44 | Black bucket for imaging |
| Krylon Industrial Acryli-Quick Lacquer Spray Paint-Ultra Flat Black | 132496 | https://www.grainger.com/product/KRYLON-INDUSTRIAL-Acryli-Quik-Spray-Paint-in-54TG76 | 2 | $8.23 | Paint for bucket |
| Sanding roll assortment | 4664A16 | https://www.mcmaster.com/#4664a16/=19720ae | 1 | $73.06 | Sanding paper for bucket |
| Tripp Lite USB 3.0 SuperSpeed Active Repeater Cable (AB M/M) 25-ft. (U328-025) | U328-025 | https://a.co/d/g4jBflW | 1 | $31.99 | For extending the Kinect USB cable if necessary |
| Kinect for Xbox one | | https://a.co/d/dJCbZDr | 1 | $83.99 | Depth camera |
| Kinect Adapter | | https://a.co/d/hYLPysv | 1 | $113.98 | Camera adaptor |
| Superlogics PC | SL-DK-AB250M-ID | https://drive.google.com/open?id=1a1ECezeF3KLXpYrAg9--HCyD-ccyqjhe | 1 | $2,505.00 | Everything you need for acquisition (PC/monitor/keyboard), be sure to get a DirectX 11 compatible video card (e.g. the NVIDIA GeForce GT 1030) |

## Table 2 | File structures

**1. File structure of the base directory at the beginning of the project**

```
<project_directory>/          ** the project directory with all depth recordings
    ├── MoSeq2-Extract-Modeling-Notebook.ipynb
    ├── MoSeq2-Analysis-Visualization-Notebook.ipynb
    ├── Flip-Classifier-Training-Notebook.ipynb
    ├── session_1/            ** the folder containing all of a single session's data
    │   ├── depth.dat         ** the recorded depth video
    │   ├── depth_ts.txt      ** txt file of the frame timestamps
    │   └── metadata.json     ** json file that contains the recording information
    ...
    └── session_n/
        ├── depth.dat
        ├── depth_ts.txt
        └── metadata.json
```

**2. File structure at the end of Procedure 2 with 1 trained model**

```
<project_directory>/
    ├── MoSeq2-Extract-Modeling-Notebook.ipynb
    ├── MoSeq2-Analysis-Visualization-Notebook.ipynb
    ├── Flip-Classifier-Training-Notebook.ipynb
    ├── progress.yaml
    ├── _pca/                  ** directory with PCA results
    ├── aggregate_results/     ** directory with all the extracted data
    ├── <base_model_path>/     ** directory with the trained model
    │   └── model.p            ** trained model
    ├── moseq2-index.yaml/     ** yaml file that contains all session metadata and group labels
    ├── config.yaml            ** yaml file that contains all parameters
    ├── session_config.yaml    ** yaml file that contains all session-specific extraction parameters
    ├── flip_classifier_k2_c57_10to13weeks.pkl    ** flip classifier
    ├── session_1/
    ...
    └── session_n/
```

**3. File structure at the end of Procedure 2 with 3 trained models**

```
<project_directory>/
    ├── MoSeq2-Extract-Modeling-Notebook.ipynb
    ├── MoSeq2-Analysis-Visualization-Notebook.ipynb
    ├── Flip-Classifier-Training-Notebook.ipynb

    ├── progress.yaml
    ├── _pca/
    ├── aggregate_results/
    ├── <base_model_path>/     ** directory with the trained model
    │   ├── model1.p           ** trained model 1
    │   ├── model2.p           ** trained model 2
    │   └── model3.p           ** trained model 3
    ├── moseq2-index.yaml/
    ├── config.yaml
    ├── session_config.yaml
    ├── flip_classifier_k2_c57_10to13weeks.pkl
    ├── session_1/
    ...
    └── session_n/
```

# Table 3 | Description of parameters in the progress file (progress.yaml)

Descriptions of parameters

| Parameters | Description |
|---|---|
| base_dir | Path to the data directory with all depth recordings. |
| config_file | Path to the config.yaml file. |
| session_config | Path to the session_config.yaml file, the file is set in Interactive Arena Mask Widget. |
| index_file | Path to the moseq2-index.yaml file. |
| train_data_dir | Path to the aggregate_results folder. |
| pca_dirname | Path to the folder containing PCA results |
| scores_filename | File name containing PCA scores. |
| scores_path | Path to the PCA scores. |
| changepoints_path | File name or path to the file containing model-free changepoints. |
| model_path | Path to the model file. |
| crowd_dir | Path to the folder that saves crowd movies. This folder should be a subdirectory of the model_session_path. |
| plot_path | Folder where most plots are saved, excluding the plots generated during the PCA step. |
| syll_info | Path to the syll_info.yaml file. The syllable labeler widget in the MoSeq2 Analysis Visualization Notebook generates a syll_info.yaml that saves the syllable names and descriptions. |
| df_info_path | Path to syllable statistics DataFrame. Contains the same information as stats_df but is saved in a different location and format. |
| base_model_path | Folder where all models and the models and the model-specific subfolders are saved. |
| model_session_path | Path to the model-specific subfolder selected for analysis. |
| snapshot | Unique ID to the existing progress. |

# Table 4 | Description of parameters in the configuration file (config.yaml)

**1. Extraction**

**1.1 Arena floor mask**

| Parameters | Description |
|---|---|
| `Floor depth range (mm)` | The depth range for detecting the floor of the arena. |
| `Floor mask dilation iterations` | The dilation iterations to expand or shrink the floor mask. |

**1.2 Extraction parameters**

| Parameters | Description |
|---|---|
| `Mouse height clip (mm)` | Minimum and maximum height to clip depth values when extracting the mouse. |
| `Number of test frames to extract` | The number of test frames to extract for preview. |
| `Display frame (index)` | The frame that is shown in the preview. |

**2. Dimensionality Reduction (Principal Component Analysis)**

| Parameters | Description |
|---|---|
| `overwrite_pca` | A boolean that specifies whether we overwrite the existing PCA results. The default is False. |
| `gaussfilter_space` | A tuple that specifies the kernel standard deviations along the horizontal and vertical directions for Gaussian filtering. The default is (1.5, 1). |
| `medfilter_space` | A parameter that specifies the kernel size for median filter on the frame. The default is 0. |
| `medfilter_time` | A parameter that specifies the kernel size for median filter across frames. The default is 0. |
| `missing_data` | A boolean that speifies whether we use missing data PCA. The default is False. |
| `missing_data_iters` | The number of times to iterate over missing data during PCA. The default is 10. |
| `recon_pcs` | The number of PCs to use for missing data reconstruction. The default is 10. |
| `dask_port` | The port to which Dask send the Diagnostics Dashboard to. The default is 8787. |
| `nworkers` | The number of workers for computing PCA. The default is 1. |

**3. Modeling**

**3.1 Most Relevant Model Parameters**

| Parameters | Description |
|---|---|
| `checkpoint_freq` | A parameter that sets the model saving freqency (in interations). The default is -1 and no checkpoint is saved. |
| `use_checkpoint` | A Boolean that indicates whether the model training is resumed from latest saved checkpoint. The default is False. |
| `npcs` | The number of PCs being used to represent the pose dynamics and they are the observations to train the AR-HMM. The default is 10. |
| `max_states` | The number of maximum states (i.e., syllables) the AR-HMM will try to learn. The default is 100. |
| `robust` | A boolean that specifies whether the noise of the syllables in the AR-HMM is sampled from a t-distribution. The default is True. |
| `separate_trans` | A boolean that specifies whether the groups are modeled with separate transition matrices. The default is False. |
| `num_iter` | A parameter that sets the number of the iterations to train the model (the number of iterations of Gibbs sampling). |
| `kappa` | A parameter that specifies the kappa setting in AR-HMM. The default is None, which equals the number of total frames. |
| `cluster_type` | A parameter that specifies the platform the modeling process runs on. The default is local. |
| `ncpus` | A parameter that specifies the number of cpus used in the model training. The default is 0. |

**3.2 Advanced model parameters - Kappa scan parameters**

CAUTION: Effective when `config_data['kappa'] == 'scan'`

| Parameters | Description |
|---|---|
| `scan_scale` | A parameter that specifies the scale to scan a range of kappa values. The default is log. |
| `min_kappa` | A parameter that specifies the minimum kappa value. The default is None, which equals 0.01 multiply by the total number of frames. |
| `max_kappa` | A parameter that specifies the maximum kappa value. The default is None, which equals 100 multiply by the total number of frames. |
| `n_models` | A parameter that specifies the total number of models to scan through. The default is 15. |
| `out_script` | The name of the file that contains all the commands for running kappa scan. |
| `run_cmd` | A Boolean that indicates whether the kappa scan script is directly run from the Jupyter notebook. The default is True. When the cell is running, the model training progress is displayed in Terminal. |

**3.3 Advanced model parameters - hold out parameters**

| Parameters | Description |
|---|---|
| `hold_out` | A boolean that specifies whether a subset of data is held out during the training process. The default is False. |
| `nfolds` | A parameter that specifies the number of folds to hold out during training if config_data['hold_out'] = True. The default is 5. |
| `cluster_type` | The compute environment the model training is running on. The default is local. |
| `ncpus` | The number of CPU cores used in the model training. The default is 0, which means all available CPU cores are used. |

# Table 5 | Dataframe column descriptions

**1. moseq_df column descriptions**

| column name | description | unit |
|---|---|---|
| angle | the orientation of the mouse body | radians |
| area_mm | the area of the mouse | mm$^2$ |
| area_px | the area of the mouse | pixels |
| centroid_x_mm | center of the mouse (x coordinate) | mm |
| centroid_x_px | center of the mouse (x coordinate) | pixels |
| centroid_y_mm | center of the mouse (y coordinate) | mm |
| centroid_y_px | center of the mouse (y coordinate) | pixels |
| height_ave_mm | average height across the entire visible mouse | mm |
| length_mm | mouse length measured roughly across the spine | mm |
| length_px | mouse length measured roughly across the spine | pixels |
| velocity_2d_mm | mouse 2D velocity (x,y velocity) | mm/frame |
| velocity_2d_px | mouse 2D velocity (x,y velocity) | pixels/frame |
| velocity_3d_mm | mouse 3D velocity (x,y,z velocity) | mm/frame |
| velocity_3d_px | mouse 3D velocity (x,y,z velocity) | pixels/frame |
| velocity_theta | direction/angle of the velocity vector | radians |
| width_mm | mouse width | mm |
| width_px | mouse width | pixels |
| dist_to_center_px | distance between mouse center and arena center | pixels |
| group | the assigned experimental group | NA |
| uuid | session uuid assigned during extraction | NA |
| h5_path | extraction h5 file path | NA |
| timestamps | frame timestamp | seconds |
| frame index | index of the frame in the recording | NA |
| SessionName | name of the session pulled from the metadata.json file | NA |
| SubjectName | name of the subject/mouse pulled from the metadata.json file | NA |
| StartTime | time of day the session recording started | NA |
| labels (original) | original syllable labels used by the AR-HMM | NA |
| labels (usage sort) | syllable label sorted by usage (low numbers = most used; high numbers = rarely used) | NA |
| labels (frames sort) | syllable label sorted by frames (low numbers = most used; high numbers = rarely used) | NA |
| onset | indicates the onset of a syllable (1/True = start of syllable) | NA |
| syllable index | syllable index | NA |

**2. stats_df column descriptions (Suffix indicates the statistics)**

| column name | description | unit |
|---|---|---|
| angle | the orientation of the mouse body | radians |
| area_mm | the area of the mouse | Mm$^2$ |
| area_px | the area of the mouse | pixels |
| centroid_x_mm | center of the mouse (x coordinate) | mm |
| centroid_x_px | center of the mouse (x coordinate) | pixels |
| centroid_y_mm | center of the mouse (y coordinate) | mm |
| centroid_y_px | center of the mouse (y coordinate) | pixels |
| height_ave_mm | average height across the entire visible mouse | mm |
| length_mm | mouse length measured roughly across the spine | mm |
| length_px | mouse length measured roughly across the spine | pixels |
| velocity_2d_mm | mouse 2D velocity (x,y velocity) | mm/frame |
| velocity_2d_px | mouse 2D velocity (x,y velocity) | pixels/frame |
| velocity_3d_mm | mouse 3D velocity (x,y,z velocity) | mm/frame |
| velocity_3d_px | mouse 3D velocity (x,y,z velocity) | pixels/frame |
| velocity_theta | direction/angle of the velocity vector | radians |
| width_mm | mouse width | mm |
| width_px | mouse width | pixels |
| dist_to_center_px | distance between mouse center and arena center | pixels |
| timestamps | averge frame timestamp in the syllable | seconds |
| frame index | average index of the frames in the syllable | NA |
| usage | the probability a syllable is used | NA |
| duration | average syllable duration | seconds |
| syllable key | sorting of the syllables (see above) | NA |
| syllable | syllable label | NA |

## Table 6 | MoSeq command summary

Summary of commands

| Operation | Command |
|---|---|
| Generate progress.yaml file and track the progress of the pipeline. | `restore_progress_vars(progress_filepath, init=True, overwrite=False)` |
| Generate config.yaml file. | `generate_config_command(config_filepath)` |
| Download flip classifier for the extraction process. | `download_flip_command(progress_paths['base_dir'], config_filepath, selection=selection)` |
| Launch the interactive arena mask widget. | `ArenaMaskWidget(progress_paths['base_dir'], progress_paths['config_file'], progress_paths['session_config'], skip_extracted=skip_extracted)` |
| Extract all the un-extracted sessions. | `extract_found_sessions(progress_paths['base_dir'], progress_paths['config_file'], extensions, extract_all=extract_all, skip_extracted=skip_extracted)` |
| Aggregate all extracted results to one folder. | `aggregate_extract_results_command(progress_paths['base_dir'], recording_format, aggregate_results_dirname)` |
| Launch the interactive tool to add group labels to the sessions. | `interactive_group_setting(progress_paths['index_file'])` |
| Train PCA for dimensionality reduction. | `train_pca_command(progress_paths, pca_dirname, pca_filename)` |
| Applying the computed PCs onto the extracted data to output dimensionality reduced data points. | `apply_pca_command(progress_paths, scores_filename)` |
| Compute model-free changepoints of the data. | `compute_changepoints_command(progress_paths['train_data_dir'], progress_paths, changepoints_filename)` |
| Train a robust AR-HMM model. | `learn_model_command(progress_paths)` |
| Set up model specific folders for analysis. | `setup_model_folders(progress_paths)` |
| Get the model that best fits the model-free changepoint, given the objective. The command supports comparison concerning two objectives: 'duration', 'jsd' and 'median_loglikelihood'.<br>• `objective='duration'`: the function returns the model where the median syllable duration best matches that of the model-free changepoints.<br>• `objective='jsd'`: the function returns the model where the syllable duration distribution best matches that of the model-free changepoints.<br>• `objective='median_loglikelihood'`: find the model that has the median loglikehood among all the trained model(s). | `get_best_fit_model(progress_paths, output_file, plot_all=True, objective=objective)` |
| Generate a DataFrame of scalar values computed during the extraction step aligned to the model syllable labels. | `scalars_to_dataframe(sorted_index, model_path=progress_paths['model_path'])` |
| Generate a DataFrame of the statistics (i.e. min, max, mea and standard deviation) of scalar values associated with each syllable. `count` parameter in the function supports `'usage'` or `'frames'`.<br>• `count='usage'`: The syllables are counted by run-length encoding (RLE) in the DataFrame.<br>• `count='frames'`: The syllables are counted by the number of frames in the DataFrame.<br>For example, for 00000 11110 11111, when `count='usage'`, syllable 0 will be 0.5 and syllable 1 will be 0.5 because RLE for the sequence would be 0101. When `count='frames'`, syllable 0 will be 0.4 and syllable 1 will be 0.6. | `compute_behavioral_statistics(scalar_df, count=count, groupby=groupby, usage_normalization=usage_normalization, syllable_key=syllable_key)` |
| Generate a DataFrame for plotting MoSeq Fingerprint and plot the Fingerprint.<br>The n_bins parameter is an integer that controls the number of bins to bin the scalar values. `preprocessor` parameter takes in a `sklearn.preprocessing` object to scale the values by session. | `create_fingerprint_dataframe(scalar_df, mean_df, n_bins=n_bins, range_type=range_type)`<br>`plotting_fingerprint(summary, progress_paths['plot_path'], range_dict, preprocessor=preprocessor)` |
| Generate syllable crowd movies and launch the interactive tool to label syllables based on crowd movies. | `label_syllables(progress_paths, max_syllables=max_syllables, n_explained=explained_variance, select_median_duration_instances=select_median_duration_instances, max_examples=max_examples)` |
| Launch the interactive tool to plot syllable statistics. | `interactive_syllable_stats(progress_paths, max_syllable=max_syllables, load_parquet=True)` |
| Launch the interactive tool to display interactive syllable transitions. | `interactive_transition_graph(progress_paths, max_syllables=max_syllables, plot_vertically=True, load_parquet=True)` |

## Table 7 | Troubleshooting table

**1. Acquisition**

| Problem | Possible reason(s) | Solution |
|---|---|---|
| Kinect2 camera is not recognized by kinect2-nidaq.exe | Camera cables are not plugged in correctly. Windows applications don't have permission to use the device. Other issues. | Check all cables and ensure they are all connected properly. When all cables are connected properly, the device should be detected by the Device Manager. Grant permissions to the device when it is used the first time. Run the diagnostics tool in Kinect for Windows SDK to find other causes for the issue. |
| Unable to run kinect2-nidaq.exe | The versions of the required software are not correct. | Check and ensure the software versions match the following:<br>- kinect2-nidaq version 0.2.3<br>- Kinect for Windows Runtime 2.0<br>- NI-DAQmx version 17.6 (with included support for .NET Frameworks 4.0, 4.5, and 4.5.1) |
| kinect2-nidaq.exe crashes immediately running. | The GPU is not Direct11 compatible. | Ensure the acquisition computer has a discrete GPU that is DirectX11 compatible. kinect2-nidaq.exe doesn't work on devices that only contains an onboard graphics card. |
| The mouse in the depth video has an anomaly patch on the back. | The camera height is too low so part of the top of the mouse is cropped off. | Measure the camera height and ensure the camera is 67.3 cm above the bottom of the arena. |
| The number of dropped frames is abnormally high. | There are other processes running that are using CPU resources. The storage space on the computer is running low. | Close other applications on computer and keep only the kinect2-nidaq.exe running. Transfer the data out of the computer to clear up storage space. |

**2. Installation**

| Problem | Possible reason(s) | Solution |
|---|---|---|
| 404 Error on GitHub. | Access to MoSeq repositories is not granted. | Email MoSeq@hms.harvard.edu to request access and accept the GitHub invitation. |
| 403 Error on GitHub. | Scopes of the access for personal tokens are not set correctly. | Check the checkbox next to repo in the "Select scopes" section when creating a new personal access token for GitHub. |
| Unable to see "`dattalab`" in REMOTE REPOSITORIES in Docker Desktop | Access to MoSeq Docker Hub repository is not granted. | Email MoSeq@hms.harvard.edu to request access. |
| Running `git clone -b release https://github.com/dattalab/moseq2-app.git` shows `fatal: Authentication failed error` after inputting GitHub username and password. | Personal token should replace the password and the GitHub credential manager is not set up. | Follow the official documentation to generate a personal token to use as your password and set up Git credential storage following this documentation. |
| Unable to successfully install autoregressive, or moseq2-model while running the `./scripts/install_moseq2_app.sh` command. | The versions of the gcc and g++ are not compatible. | Check the gcc and g++ version running:<br>`>>> which gcc`<br>`>>> which g++`<br><br>Windows and Linux users should have gcc and g++ higher than 6 and MacOS users should have gcc-7 and g++-7.<br><br>Export the gcc and g++ environment variable after the compatible versions are installed. |

**(Continued)**

## Table 7 | Troubleshooting table (cont.)

**3. MoSeq2 package suite**

| Problem | Possible reason(s) | Solution |
|---|---|---|
| The detected arena mask doesn't cover the whole arena. | The depth range is not set properly for the recording setup. | Adjust the Floor depth range slider and compute the arena mask again until the correct area is detected. |
| Videos fail to display in the interactive widgets. | The browser or IDE is not compatible with JavaScript and HTML widget. | Use Chrome for the Jupyter notebooks. |
| Explained variance of over 90% requires over 15 Principal Components (PCs). | The acquired data is too noisy and (or) the bucket is too reflective. | Examine the extracted data further and check if the bucket needs more sanding and painting. |
| The AR-HMM modeling step shows `numpy.linalg.LinAlgError: Matrix is not positive definite.` | The average velocity of some sessions is almost 0, indicating the mice are almost stationary most of the time. | Examine the extracted data further and/or exclude the sessions in the modeling step. |
| Compute scalar_df shows `AssertionError: The pca_path variable in the index file is not pointing to the correct file.` | The `pca_path` in moseq2-index.yaml file is empty, likely because moseq2-index.yaml is re-generated after the PCA steps. | Add the path to the pca_scores.h5 file to `pca_path` in moseq2-index.yaml. |
| The interactive syllable statistics graphing takes a long time to refresh. | There are too many groups to plot. | Consider setting fewer groups or use the non-interactive CLI alternatives. |
| Running the cell(s) in the Analysis and Visualization notebook shows `NameError: name 'max_syllables' is not defined` | Interactive Syllable Labelling Tool is not run and `max_syllables` is not defined by the user. | Run the Interactive Syllable Labelling Tool to set max_syllables based on an explained variance percentage or specify max_syllables variable to a specific integer that represent the desired number of syllables to include. |

**Author Contributions**

SL, WFG, CW and AZ wrote the codebase and the protocol, SCJ, EMR and SRD tested the protocol, SL, WFG and SRD wrote and edited the paper.

**Acknowledgements**

SRD is supported by SFARI, the Simons Collaboration on the Global Brain, the Simons Collaboration for Plasticity and the Aging Brain, and by NIH grants U24NS109520, U19NS113201, R01NS114020, and RF1AG073625. WFG is supported by NIH grant F31NS113385, and CW is supported by a postdoctoral fellowship from the Jane Coffin Childs foundation.

**Competing Interests**

SRD sits on the scientific advisory boards of Neumora, Inc. and Gilgamesh Therapeutics, which have licensed the MoSeq technology.